\documentclass[reprint,amsmath,amssymb]{revtex4}

\usepackage{graphicx}
\usepackage{amsmath}
\usepackage{dcolumn}
\usepackage{bm}
\usepackage{xcolor}
\usepackage{algorithm} 
\usepackage{algpseudocode} 
\usepackage[utf8x]{inputenc}
\begin{document}

\title{WAND-PIC: an accelerated three-dimensional quasi-static particle-in-cell code}

\author{Tianhong Wang$^1$, Vladimir  Khudik$^2$, Jihoon Kim$^1$, and Gennady Shvets$^1$}\affiliation{$^1$School of Applied and Engineering Physics, Cornell University, Ithaca, New York 14850, USA. \\ $^2$Department of Physics and Institute for Fusion Studies, The University of Texas at Austin, Austin, Texas 78712, USA.}

\date{\today}
\begin{abstract}
We introduce a quasi-static particle-in-cell (PIC) code -- WAND-PIC -- which does not suffer from some of the common limitations of many quasi-static PICs, such as the need for a predictor-corrector method in solving electromagnetic fields.  We derive the field equations under quasi-static (QS) approximation and find the explicit form of the "time" derivative of the transverse plasma current. After that, equations for the magnetic fields can be solved exactly without using the predictor-corrector method.  Algorithm design and code structure are thus greatly simplified. With the help of explicit quasi-static equations and our adaptive step size, plasma bubbles driven by the large beam charges can be simulated efficiently without suffering from the numerical instabilities associated with the predictor-corrector method. In addition, WAND-PIC is able to simulate the sophisticated interactions between high-frequency laser fields and beam particles through the method of sub-cycling. Comparisons between the WAND-PIC and a first-principle full PIC code (VLPL) are presented. WAND-PIC is open-source~\cite{WAND_PIC}, fully three-dimensional, and parallelized with the in-house multigrid solver. Scalability, time complexity, and parallelization efficiency up to thousands of cores are also discussed in this work.
\end{abstract}

\maketitle

\section{Introduction}
Plasma-based accelerators represent one of the most exciting concepts in high-gradient particle acceleration. Plasmas with density $n_0$ can sustain a high accelerating gradient $E_{\parallel} \sim \sqrt{n_0/10^{18}{\rm cm^{-3}}} [{\rm GV/cm}]$, thereby enabling compact particle accelerators that are much smaller than the present-day conventional accelerators. Accelerating structures are excited either by ultra-intense laser pulses for a laser wakefield accelerator (LWFA)~\cite{LWFA_1,LWFA_2,LWFA_3}, or by relativistic electron bunches for a plasma wakefield accelerator (PWFA)~\cite{PWFA_1}. GeV-level electron accelerations have been demonstrated in recent experiments for both LWFA~\cite{GeV_0,GeV_1,GeV_2,GeV_3,GeV_2b} and PWFA~\cite{PWFA_GeV_1,PWFA_GeV_2,FACET_II_PWFA,FACET_II_PWFA_II}.

In LWFA, the energy gain of the witness bunch over a dephasing distance $L_d\propto n_0^{-3/2}$~\cite{Tajima_1979,Joshi_1984,Lu_GeV} can be estimated as $\Delta W_{\rm LWFA} = E_{\parallel} L_d \propto n_0^{-1}$ which favors the use of low density plasma and long propagation distance.  Plasma densities $n_0 \sim 10^{17}{\rm cm^{-3}}$ are employed in recent experiments~\cite{GeV_2b} with energy gain reaching $8$GeV over an acceleration distance of $20$cm. In PWFA, the energy gain of the witness bunch is largely governed by the transformer ratio~\cite{TR_PisinChen_PRL86} and the initial energy of the driver beam $\gamma_b mc^2$, where $\gamma_b$ is the relativistic factor of the driver particles. The maximum energy gain of an accelerated (witness) electron beam is limited to $\Delta W_{\rm PWFA} = 2\gamma_b mc^2$. Energy doubling of 42 GeV electrons in a meter-scale plasma has been demonstrated \cite{blumenfeld_energy_2007}. Therefore, either in LWFA or PWFA, increasing the energy gain would inevitably require more energy in the driver and a longer propagation distance. With the rapid development of ultra-intense multi-petawatt laser systems~\cite{Korea_4PW,mourou2011eli_whitebook,Apollon_10PW} and ultra-short, high-current compact electron beam sources~\cite{FACET-II-Design,FACET-II-2020}, one can anticipate an increasing number of LWFAs and PWFAs with per-stage lengths on the order of a meter. However, simulating meters-long propagation distances of plasma-based accelerators currently presents a computational challenge. Numerical challenges escalate even further when hundreds of meter-scale stages required for developing TeV-scale linear lepton colliders~\cite{Colliders} must be accurately modeled.

Generally speaking, there are two major approaches to simulating plasma-based accelerators: the first-principles and the reduced-description (quasi-static) particle in cell (PIC) simulations. The first-principles PIC approach~\cite{PIC_1,PIC_2} is based on explicitly solving driven Maxwell's equations for the electric and magnetic fields on a staggered Yee grid~\cite{YEE} using a finite-difference time-domain (FDTD) method. Electric currents carried by charged macroparticles are interpolated onto the grid; they serve as driving source terms for Maxwell's equations. In turn, the macroparticles are advanced in time using the calculated electromagnetic fields. A crucial constraint of the first-principles PICs is the Courant–Friedrichs–Lewy (CFL) condition~\cite{CFL_Condition} that must be satisfied to avoid numerical instabilities: the product of the time step $\Delta t$ and the speed of light $c$ must be smaller than the spatial step size, which must be chosen to be much smaller than the smallest length scale $\Delta L_{\min}$ in the simulation domain.

For example, the smallest length scale for simulating an LWFA is typically determined by the laser wavelength: $\Delta L_{\rm min}^{\rm LWFA}=\lambda_0$. Therefore a time step size $\delta t<<\omega_0^{-1}$ is required, where $\omega_0=2\pi c/\lambda_0$ is the laser frequency. Therefore, the CFL condition imposes severe constraints on faithful 3D simulations of LWFAs: more than $10^4$ core-hours are required to simulate a propagation distance of just a few millimeters assuming that $\lambda_L \sim 0.8\mu m$. Likewise, for a PWFA operating in the strongly-nonlinear regime characterized by a complete blowout of plasma electrons from the path of the driver bunch, the smallest length scale normally equals the sharpness of the nonlinear wakefield: $\Delta L_{\rm min}^{\rm PWFA}<<c/\omega_p$, where $\omega_p = \sqrt{4\pi e^2n_0/m}$ is the plasma frequency, $-e$ and $m$ are the electron charge and mass, respectively.  It is worthwhile to mention that $\Delta L_{\rm min}^{\rm PWFA}$ approaches zero when wave-breaking happens in a cold plasma~\cite{WB_JETP_1956,WB_Dawson_1959}. This density singularity can be prevented by adding finite temperature~\cite{WB_Warm_Coffey,WB_Warm_Schroeder,WB_Warm_Schroeder2}. However, for typical PWFA parameters, the $\Delta L_{\rm min}^{\rm PWFA}$ remains small.

The quasi-static approach emerges from a simple observation that in many realistic scenarios, the characteristic evolution times $\Delta T_{\rm dr}$ of the drivers are much longer than their corresponding durations: $\Delta T_{\rm LWFA} \sim \omega_0/\omega_p^2 \gg \tau_{\rm LWFA}$ and $\Delta T_{\rm PWFA} \sim \sqrt{2\gamma}/\omega_p \gg \tau_{\rm PWFA}$ for the laser and beam drivers, respectively. Therefore, great savings of computational time could be achieved if these two distinct time scales could be explicitly separated in a code. This is done using the so-called quasi-static approximation (QSA) proposed by Sprangle {\it et al}~\cite{Sprangle_1990} and originally implemented in a PIC code by Mora {\it et al}~\cite{Mora_1996,Mora_1997} and Whittum~\cite{Whittum_1997}. The QSA assumes that the envelope of the driver is "frozen" during the time when cold plasma electrons are passed over by the driver, and enables a time step size $\Delta t \sim \Delta T_{\rm dr}$ which is much longer than the one imposed on any first-principles PIC code by the CFL condition. Indeed, several quasi-static PIC codes, such as WAKE~\cite{Mora_1996,Mora_1997}, LCODE~\cite{lotov_2003,lotov_2004}, QuickPIC~\cite{QUICK_2006,QUICK_2009,QUICK_2013}, and HiPACE~\cite{HiPACE_2014}, have demonstrated computational time reductions of over two orders of magnitude over their first-principles counterparts. The WAKE and LCODE are two-dimensional (2D) codes with cartesian or cylindrical geometry, and QuickPIC and HiPACE are fully-3D and fully parallelized.

However, such impressive computational time savings are accompanied by several additional numerical complications. Unlike the FDTD method, where the fields are naturally discretized in time and space on a Yee grid and can be locally updated from the previous time step, a quasi-static PIC code must regenerate new wakefields at every time step because the calculations of the wakefields and drivers are decoupled. Yet a bigger challenge is that in some equations of wakefields under QSA, the source contains a time derivative of the transverse current, which is not explicitly expressed~\cite{Mora_1997}. Specifically, the equation for the transverse magnetic field of the wake is not expressed in a closed and explicit form and is usually solved using the so-called "predictor-corrector" method~\cite{Predictor-Corrector}. To our knowledge, this approach is taken in all the widely-used quasi-static codes (WAKE, LCODE, QuickPIC, HiPACE). While an improved iteration loop~\cite{QUICK_2013} has been developed to improve the stability and convergence of the predictor-corrector method, current quasi-static codes are still challenged in simulating extremely nonlinear wakefields that exist, for example, inside a fully-evacuated plasma bubble that develops in the full blowout regime~\cite{rosen_pra91,pukhov_AppPhysB02}. The reason is that the predictor-corrector method often fails to converge at the back of the large bubble, where plasma electrons are highly relativistic and the wakefields are sharp~\cite{lotov_2003}. This long-standing issue of quasi-static codes, which we address and resolve in this work, has prevented rapid and accurate investigations of the accelerating structures driven by the most powerful beam and laser drivers.

In this work, we describe a fully 3D massively-parallel quasi-static code, WAND-PIC (\textbf{W}akefield \textbf{A}cceleratio\textbf{N} and \textbf{D}irect laser acceleration), which does not use the predictor-corrector method. New WAND-PIC is a major update to the non-parallelized azimuthally-symmetric version of WAND-PIC~\cite{My_Driver_2017} that cannot capture important beam- and laser-plasma phenomena such as hosing~\cite{Hosing_1,Hosing_2}. A newly-derived set of quasi-static field equations with fully-explicit source terms are applied to simulating large plasma bubbles. The massively parallel nature of WAND-PIC can now be run on distributed computer clusters comprising thousands of computational cores. Adaptive longitudinal step size refinement and nonuniform transverse grids enable efficient simulations of large evolving plasma bubbles containing regions with different spatial scales. For the first time, accurate simulation of the direct laser acceleration (DLA) of electrons with a quasi-static code is enabled by implementing sub-cycling in WAND-PIC.

The rest of the paper is organized as follows. First, we summarize the advanced features in WAND-PIC in Section~\ref{sec_Summ}. The quasi-static equations for fields and particles are described in \ref{sec_ALL_EQS}. The implementation of different drivers and their interactions are presented in Section~\ref{sec_laserfields}. In Section~\ref{sec_compare}, we compare the results of the WAND-PIC and the first-principles 3D PIC code VLPL~\cite{Pukhov_code}. Then in Section~\ref{CEP_PUB}, we present a simulation example where a DLA-induced phase-dependent bubble undulation is modeled.  The algorithm design and parallelization efficiency are discussed in Section~\ref{sec_benchmark}, followed by a brief discussion of the future code development and conclusions.

%

\section{Summary of the architecture and key features of WAND-PIC}\label{sec_Summ}
\begin{figure}[htp!]
\centering
\includegraphics[width=0.8\columnwidth]{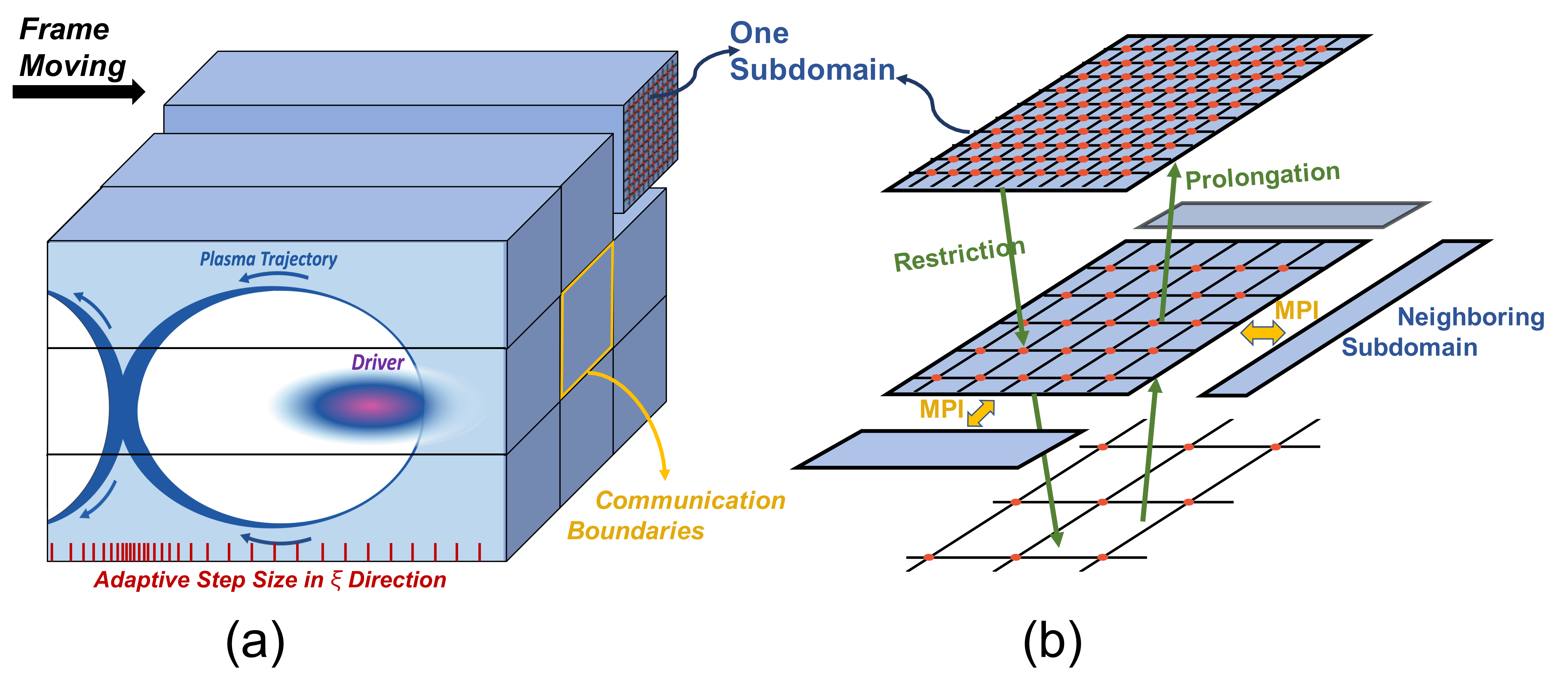}\\
\caption{(a) The 3D simulation domain in WAND-PIC. The 2D transverse plane is partitioned into square subdomains. In the longitudinal direction, the adaptive step size is shown by the red ticks which are denser at the back of the bubble. (b) The multigrid approach: shown is a hierarchy of grids in one subdomain with different mesh sizes.}
\label{Fig1}
\end{figure}
In this section, we briefly introduce the four advanced features of WAND-PIC.
\begin{enumerate}
\item Explicit source terms for all wakefield equations\\
In quasi-static codes, the particles' distribution and the fields depend on $t$ and $z$ only through a "time-like" variable $\xi=ct-z$. Therefore, the full 3D domain consists of a stack of 2D planes (slices) propagated in $\xi$ with variable steps $d\xi$. Electromagnetic fields are computed at every 2D $\xi$-slice as explained in Sec.~\ref{sec_ALL_EQS}. Macroparticles are pushed from one slice to the next with the same variable $\xi$-step because they propagate strictly in the positive $\xi$-direction: $d\xi/dt=c-v_z>0$. Therefore, when solving for the fields at a given $\xi$ step, only information only at the current ($\xi$) step and the previous ($\xi-d\xi$) steps are known. Ideally, the source terms of all field equations are explicitly expressed, i.e. no "future" information at ($\xi+d\xi$) step is needed. Detailed derivations of Maxwell equations under the QSA are available~\cite{Mora_1997,lotov_2003,QUICK_2013,HiPACE_2014}, and the results are largely similar: the equations for at least some electromagnetic field components are given in the implicit form. For example, the source term $\partial \bf{j}_\perp/\partial\xi$ is implicitly expressed in the equations for either the transverse components of the magnetic (QuickPIC and HiPACE) or electric (LCODE) fields $\bf{B_\perp}$ and $\bf{E_\perp}$, or for the wakefield potential $\psi$ (WAKE). In contrast, the same source term is explicitly expressed in WAND-PIC. Therefore, all field equations are expressed in a closed "static" form (since $\xi$ functions as a time-like variable) and can be solved just once for each $\xi$-slice using any optimized Poisson solver. No predictor-corrector iterations of the field solver are necessary: see Sec.~\ref{sec_ALL_EQS} for more details.

\item Adaptive step size refinement\\
With an increasing need for simulating plasma wakefield driven by the tightly-focused laser pulses with peak powers exceeding $P_L \sim 1 {\rm PW}$ and beam drivers with peak currents exceeding $I_b \sim 10 {\rm kA}$, it is imperative to ensure that the nonlinearity of wakefield/bubble structure is accurately modeled over tens of centimeters (or even meters for a PWFA). In WAND-PIC, we have implemented the technique of adaptive step size refinement in the longitudinal ($\xi$) direction to handle the steep wakefield structure at the back of the bubble (see Fig.~\ref{Fig2}(b) for a typical example). As shown in Fig.~\ref{Fig1}, the longitudinal step sizes $d\xi$ (red ticks) are automatically adjusted based on the speeds of plasma trajectories: a finer mesh is used near the back of the bubble, where plasma electrons are the fastest. More details and simulation examples can be found in Sec.~\ref{sec_beam_driver}.

\item Full description of driver particles in high-frequency laser fields\\
A delicate situation can occur when two co-propagating drivers -- a laser pulse and an electron bunch -- are employed to drive a plasma wake. Such a situation can occur, for example, in a Laser-pulse and Electron-bunch Plasma-wakefield Accelerator~\cite{My_LEPA_2020}. Unlike the ambient plasma electrons, those comprising the driver bunch cannot be described within either the QSA framework or the ponderomotive (phase-independent) approximation. Therefore, we improved the modeling of the driver bunch particles by incorporating the high (optical) frequency laser fields into their equations of motion. This can be done because even within the QSA description, both the envelope and the phase of the laser pulse are calculated at each time step of the laser advance.  Using the sub-cycling method~\cite{DLA_require_Alex}, the particles are advanced with a sufficiently small time step which is much smaller than that of the laser advances. More details and specific examples are presented in Sections \ref{sec_laserfields} and \ref{sec_DLA}.

\item Parallel geometric multigrid solver on nonuniform grids\\
In quasi-static code, fields are solved at every 2D $\xi = {\rm const}$ slice, and particles are advanced in the positive $\xi$-direction from one $\xi$ slice to the next. For the fields, a 2D Poisson solver that uses grid-interpolated particle densities and currents as sources for computing the electromagnetic fields of the plasma wake must be utilized. In order to make WAND-PIC compatible with the present-day high-performance computing systems, we parallelized the transverse space ${\bf{r}}\equiv(x,~y)$ by applying a square partitioning on it. As shown in Fig.~\ref{Fig1}, the transverse plane is divided into square-shaped sub-domains. An in-house parallel multigrid(MG) solver~\cite{Multigrid_1,Multigrid_2} is developed alongside WAND-PIC. This geometric MG solver naturally accommodates the 2D partitioning of the computational domain; it employs iterative relaxation of the solution on a hierarchy of grids with different grid sizes. As a result, residual errors with different spatial scales are effectively smoothed out at different layers. This MG solver is also compatible with nonuniform transverse grids, i.e., WAND-PIC can use finer grids at the area of interest to improve the modeling accuracy and reduce the computational cost at the same time. For example, finer grids can be deployed at the bubble region, and coarse grids are used for the plasma outside. Details of the implementation of nonuniform grids are discussed in Sec.~\ref{sec_Nonuniform} and performance evaluation of the parallel MG solver is presented in Sec.~\ref{sec_benchmark}.

\end{enumerate}

\section{Quasi-static equations for wakefields and plasma particles}\label{sec_ALL_EQS}
The quasi-static equations used in WAND-PIC were originally derived in~\cite{My_Driver_2017} for azimuthally-symmetric wakes. General 3D equations that do not make any symmetry assumptions are very similar, and their derivation is briefly described below.  In what following, we use dimensionless units normalizing time  to $\omega_p^{-1}$, length to $k_p^{-1}$, and velocities to  $c$. We also normalize the electron kinetic momentum ${\bf{p}}$ to $mc$, the fields ${\bf{E}}$ and ${\bf{B}}$ to $mc\omega_p/e$, the potentials $\phi$ and {\bf{A}} to $mc^2/e$, the plasma density to $n_0$, and the current density  ${\bf{j}}$  to $-en_0c$.

We start with deriving, under the QSA, the time-evolution equation for the electron distribution function $f_e(t,{\bf{R}},{\bf{P}})$ in the three-dimensional phase space $({\bf{R}},{\bf{P}})$ , where ${\bf{R}}=({\bf r},z)$ and $\bf{P}$ are the 3D position and momentum, respectively. In general, $f_e$ traces the following phase space trajectory:
\begin{eqnarray}
&&\frac{\partial {f}_e}{\partial t} +\frac{\partial H}{\partial {\bf{P}}}\cdot\frac{\partial f_e}{\partial {\bf{R}}}-\frac{\partial H}{\partial {\bf{R}}}\cdot\frac{\partial f_e}{\partial {\bf{P}}}=0,\label{eq:Vlasov_3D}
\end{eqnarray}
where $H=[1+({\bf{P}}+{\bf{A}})^2]^{1/2}-\phi$ is the relativistic Hamiltonian normalized to the electron rest energy $mc^2$. The trajectory of an individual plasma electron is determined by the Hamilton's equations of motion:
\begin{eqnarray}
&&\frac{d{\bf{P}}}{dt}=-\frac{\partial H}{\partial {\bf{R}}}, \quad \frac{d{\bf{R}}}{dt}=\frac{\partial H}{\partial {\bf{P}}} = \frac{\bf{p}}{\gamma},
\label{eq:EoM_3D}
\end{eqnarray}
where $\bf{p} = {\bf{P}} + \bf{A}$ and $\gamma = \sqrt{1 + \bf{p}^2}$ are, respectively, the kinematic momentum (normalized to $mc$) and the relativistic factor of a plasma electron.

Equations (\ref{eq:Vlasov_3D},\ref{eq:EoM_3D}) can now be simplified because, under the QSA, all electromagnetic fields are assumed to be dependent on time $t$ and coordinate $z$ through their combination $\xi=t-z$, where the speed of light $c$ is normalized to $1$: $\phi \equiv \phi({\bf r},\xi)$ and ${\bf A} \equiv {\bf A}({\bf r},\xi)$. This crucial simplification is based on the assumption of a relativistic plasma wake, i.e. it assumes that all plasma fields are excited by a driver -- laser pulse, electron bunch, or both -- that are moving through the plasma with highly-relativistic velocity $v_{\rm dr}(t)$ (in general, time-dependent) satisfying $c - v_{\rm dr} \approx c/(2\gamma_{\rm dr}^2)$, where $\gamma_{\rm dr}\gg 1$. Naturally, this limits the applicability of the QSA to the plasma wakes excited by an ultra-relativistic charged bunch (in which case, $\gamma_{\rm dr} = \gamma_b$, where $\gamma_b mc^2$ is the energy of the bunch particles), or by a laser pulse propagating through tenuous plasma (in which case $\gamma_{\rm dr} \sim \omega_0/\omega_p$). However, as we demonstrate in Section~\ref{sec_beam_driver} through comparisons with full-PIC simulations, the structure of the wakefield is quantitatively accurate when modeled within the QSA framework that takes the $\gamma_{\rm dr} \rightarrow \infty$ limit. On the other hand, the process of self-injection and subsequent trapping of plasma electrons into the plasma wake is highly sensitive to $\gamma_{\rm dr}$. While some progress has been made in modeling such processes with quasi-static codes~\cite{morshed_pop10,jain_pop15}, those are presently outside of the scope of WAND-PIC. In the rest of this Section, we separately discuss the equations of motion for plasma electrons, wakefields, and the driver (charged bunch and laser pulse).

\subsection{Description of plasma electrons motion}\label{subsec:plasma_electrons}

Under the QSA, plasma electrons dynamics can be described as a 2D motion in the $(x,y)$ plane as a function of a time-like parameter $\xi$. While the electron Hamiltonian $H$ is not conserved as a function of $\xi$ because it is a function of $\xi$-dependent scalar and vector potentials $\left( \phi, {\bf A} \right)$, it posses one integral of motion. Specifically, from $dH/dt={\partial H}/{\partial t}={\partial H}/{\partial \xi}$  and $dP_z/dt=-{\partial H}/{\partial z}={\partial H}/{\partial \xi}$, we find the following conserved quantity: $H-P_z = {\rm const}$.  Assuming that all electrons comprise a cold homogeneous plasma, i.e. $H=1$ and ${\bf{P}}=0$ for every plasma electron prior to the arrival of the driver, the integral of motion takes the form: $H-P_z-1=0$. Therefore, the electron distribution function $f_e$ can be expressed in the following form:
\begin{eqnarray}
f_e(t,{\bf{R}},{\bf{P}})=f_{*}(\xi,{\bf{r}},{\bf{P}}_{\perp})\delta(H -P_z-1),
\label{eq:substitute_A}
\end{eqnarray}
where ${\bf P} = \left( {\bf P}_{\perp}, P_z \right)$, and $f_*$ represent a distribution function of macroparticles moving in the transverse plane $(x,~y)$. Substituting Eq.~(\ref{eq:substitute_A}) into (\ref{eq:Vlasov_3D}), we find that $f_*$ satisfies the following Vlasov-like equation:
\begin{eqnarray}
&&\frac{\partial {f}_*}{\partial \xi} +\frac{\partial H_*}{\partial {\bf{P}}_{\perp}}\cdot\frac{\partial f_*}{\partial {\bf{r}}}-\frac{\partial H_*}{\partial {\bf{r}}}\cdot\frac{\partial f_*}{\partial {\bf{P}}_{\perp}}=0,\label{eq:Vlasov_2D_A}
\end{eqnarray}
where
\begin{eqnarray}
 H_*=\frac{1+({\bf{P}}_{\perp}+{\bf{A}}_{\perp})^2+(1+\psi)^2}{2(1+\psi)}-\psi-A_z\label{eq:H_2D_A}
\end{eqnarray}
is the Hamiltonian for the two-dimensional motion in the $(x,y)$-plane and $\psi=\phi-A_z$ is the wakefield potential.  The trajectory of an individual particle, as it advances in $\xi$ through the $({\bf{r}},{\bf{P}}_{\perp})$ phase space, is obtained from Eq.~(\ref{eq:EoM_3D}). Recalling that ${\bf{P}}_{\perp}={\bf{p}}_{\perp}-{\bf{A}}_{\perp}$, these equations can be recast in the following form:
\begin{eqnarray}{}
&&\frac{d}{d\xi}{\bf{r}}_{\perp} \equiv {\bf{V}}_{\perp} = \frac{1}{1+\psi}{\bf{p}}_{\perp} \label{eq:EoM_112_A} \\
&&\frac{d}{d\xi}{\bf{p}}_{\perp}=\frac{\gamma\nabla_{\perp}\psi}{1+\psi} +[ {\bf{e}}_z\times {\bf{V}}_{\perp}]B_z+[{\bf{e}}_z\times {\bf{B}}_{\perp}] - \frac{\nabla_{\perp}(|\hat{\bf{A}}_{\perp}|^2/4)}{1+\psi},\label{eq:EoM_111}
\end{eqnarray}
where  $\gamma=[{1+{\bf{p}}_{\perp}^2+(1+\psi)^2}+|\hat{\bf{A}}_{\perp}|^2/2]/{2(1+\psi)}$ is the relativistic factor, ${\bf{V}}_{\perp} \equiv {\bf{p}}_{\perp}/({1+\psi})$ is the effective particle "velocity" in the $(x,y)$-plane as it advances in $\xi$, and the last term in Eq.~(\ref{eq:EoM_111}) is the ponderomotive force produced by a laser pulse with transverse vector potential given by $\tilde{\bm{A}}_{\perp} =\hat{\bm{A}}_{\perp}\exp[-ik_0\xi]$~\cite{Mora_1996,Mora_1997}.
Note that $\psi$, $B_z$, and $\bf{B}_{\perp}$ are the only plasma wakefields that are needed to advance plasma electrons under the QSA. These fields are functions of $\left(\xi, {\bf r}_{\perp} \right)$, and are updated at each step $\xi$ as described below. The complex-valued laser envelope $\hat{\bm{A}}_{\perp}$ is advanced in time $t$ as described in Sec.~\ref{sec_beam_driver}.

\subsection{Description of plasma wakefields under the QSA}\label{subsec:plasma_wakes}
Below we describe the calculation of field quantities, such as electromagnetic fields, as well as plasma fluid quantities such as density, velocity, and pressure tensor.
Integration of  Eq.~(\ref{eq:Vlasov_2D_A}) over the transverse momenta  $P_x$ and $P_y$ yields the continuity equation:
\begin{eqnarray}
&&\frac{\partial}{\partial \xi}n_{*}=
-\nabla_{\perp}\cdot (n_*\langle{\bf{V}}\rangle),
\label{eq:continuity}
\end{eqnarray}
where  $n_{*}({\bf r},\xi) \equiv \int dP_xdP_y f_*$ and $\langle {\bf{V}}_{\perp} \rangle({\bf r},\xi) = {{n}}_{*}^{-1}\int dP_xdP_y f_{*} {\bf{V}}_{\perp}$ are the surface electron density and transverse velocity, respectively. As seen from Eq.~(\ref{eq:continuity}), the total particle number $N = \int d^2 {\bf r} n_{*}$ is $\xi$-independent. The 3D number $n_e$ and current ${\bf{j}}_{\perp}$ densities of plasma electrons can be expressed as follows:
\begin{eqnarray}
n_e=\frac{n_*\langle \gamma \rangle}{1+\psi}, \quad
{\bf{j}}_{\perp}=n_*\langle {\bf{V}}_{\perp} \rangle, 
\quad j_z=\frac{n_*\langle p_z\rangle}{1+\psi},\label{eq:Jr}
\end{eqnarray}
where $\langle p_z\rangle=\langle \gamma \rangle-\psi-1=[1+\langle {\bf{p}}_{\perp}^2\rangle-(1+\psi)^2+|\hat{\bf{A}}_{\perp}|^2/2]/(2(1+\psi)$. Note that $n_e-j_z=n_*$.

We now show that the wakefield potential $\psi$  and the fields $E_z,~B_z$  at the same slice position $\xi$ are determined only by electrons' positions and momenta at the same slice, i.e. by  $f_{*}(\xi,{\bf{r}},{\bf{P}}_{\perp})$. Under the QSA, Maxwell's equations in dimensionless variables take the following form:
\begin{eqnarray}
&&\nabla\times {\bf{E}}=-\frac{\partial}{\partial \xi}{\bf{B}},
\label{eq:ME_1}\\
&&\nabla\times {\bf{B}}=\frac{\partial}{\partial \xi}{\bf{E}}-{\bf{j}}.
\label{eq:ME_2}
\end{eqnarray}

Combined with Gauss's law $\nabla\cdot {\bf{E}}=-n_e+1$, and using the transverse Coulomb gauge ${\bf \nabla}_{\perp} \cdot {\bf A}_{\perp} = 0$~\cite{Mora_1997}, we obtain the following set of equations:
\begin{eqnarray}
&&\nabla_{\perp}^2 \psi = n_{*}-1, \label{eq:EqM_N8_AA}\\
&&\nabla_{\perp}^2 E_z=-\nabla_{\perp}\cdot{\bf{j}}_{\perp},
\label{eq:EqM_N8_A}\\
&&\nabla_{\perp}^2 B_z={\bf{e}}_z\cdot[\nabla_{\perp}\times {\bf{j}}_{\perp}], \label{eq:Mag_z_A}\\
&&\nabla_{\perp}^2 {\bf{B}}_{\perp}=-[{\bf{e}}_z\times\nabla_{\perp}j_z]
-\bigg[{\bf{e}}_z\times\frac{\partial }{\partial \xi} {\bf{j}}_{\perp}\bigg]. \label{eq:Mag_x_A}
\end{eqnarray}

We note that, while the $E_z$ field is not explicitly used in Eqs.~(\ref{eq:EqM_N8_A},\ref{eq:EqM_N8_AA}) describing plasma electrons motion, it is nevertheless important for simplifying the rhs of Eq.~(\ref{eq:Mag_x_A}). Specifically, while Eqs.~(\ref{eq:EqM_N8_AA},\ref{eq:EqM_N8_A},\ref{eq:Mag_z_A}) are {\it local} in $\xi$ (i.e. solving them only require that we calculate electrons' positions and momenta at the 2D slice of interest), Eq.~(\ref{eq:Mag_x_A}) containing a $\xi$-derivative which is not local. From a computational standpoint, accurate calculation of this term requires that we know the value of ${\bf{j}}_{\perp}$ at several $\xi$-slices. Traditionally, the $\partial {\bf{j}}_{\perp}/\partial \xi$ term has been calculated using the Predictor-Corrector approach~\cite{Predictor-Corrector}. As shown below, calculating an additional quantity $E_z$ enables us to replace Eq.~(\ref{eq:Mag_x_A}) containing a non-local source with another one that does not contain any non-local quantities.

To obtain a local (i.e. free of derivatives in $\xi$) form of Eq.~(\ref{eq:Mag_x_A}), we establish an additional relationship between the $\xi$-derivative of the transverse current and the electromagnetic fields in a manner similar to the way it was done for an azimuthally-symmetric problem~\cite{My_Driver_2017}: we multiply Eq.~(\ref{eq:Vlasov_2D_A}) by the "velocity" ${\bf{V}}_{\perp}=\partial H/\partial{\bf{P}}_{\perp}$ and integrate it over momentum. After straightforward calculations, we establish the following relativistic fluid equation for the transverse current density:

\begin{eqnarray}
\frac{\partial}{\partial\xi}{\bf{j}}_{\perp} = n_{*}\langle {\bf{a}}_{\perp} \rangle
- {\bf \nabla}_{\perp} \cdot \langle n_{*} {\bf{V}}_{\perp} {\bf{V}}_{\perp} \rangle,
\label{eq:FluxPres}
\end{eqnarray}
where ${\bf T}({\bf r}) \equiv n_{*} \langle {\bf{V}}_{\perp} {\bf{V}}_{\perp} \rangle$ is the pressure tensor, and ${\bf{a}}_{\perp} \equiv {d^2{\bf{r}}_{\perp}}/{d\xi^2}$ is the transverse relativistic acceleration:
\begin{equation}\label{eq:accel_A}
  {\bf{a}}_{\perp} = \frac{[{\bf{e}}_z\times {\bf{B}}_{\perp}]}{1+\psi} + \frac{[{\bf{e}}_z \times  {\bf{V}}_{\perp}]B_z}{(1+\psi)} + \frac{\gamma\nabla_{\perp}\psi-\nabla_{\perp}(|\hat{\bf{A}}_{\perp}|^2/4)}{(1+\psi)^2} - \frac{{\bf{V}}_{\perp}}{1+\psi} \bigg(E_z + {\bf{V}}_{\perp} \cdot \nabla_{\perp} \psi \bigg).
\end{equation}

Note that, while $E_z$ does not explicitly enter Eq.~(\ref{eq:EoM_111}) for the transverse kinematic momentum ${\bf p}_{\perp}$, it enters the expression for the transverse relativistic acceleration ${\bf{a}}_{\perp}$. For brevity, we have also suppressed the implicit dependence of ${\bf T}$, $n_{*}$, and $j_z$, and $\langle {\bf V}_{\perp} \rangle$ on the slice index $\xi$.

After substituting ${\partial {\bf{j}}_{\perp}}/{\partial\xi}$ from Eq.~(\ref{eq:FluxPres}) into Eq.~(\ref{eq:Mag_x_A}), we obtain Helmholtz-like inhomogeneous equation for the transverse magnetic field:
\begin{eqnarray}
&&\nabla_{\perp}^2 {\bf{B}}_{\perp} - \frac{n_{*}}{1+\psi}{\bf{B}}_{\perp} = - [{\bf{e}}_z \times{\bf{S}}],
\label{eq:Main_Eq_A}
\end{eqnarray}
where the rhs contains a source ${\bf S}$ given by
\begin{equation}\label{eq:FluxCurr1_A}
  {\bf{S}} = {\bf \nabla}_{\perp} j_z - {\bf \nabla}_{\perp} \cdot \left( n_{*} \langle {\bf{V}}_{\perp} {\bf{V}}_{\perp} \rangle \right) + \frac{n_{*}{\bf{e}}_z\times \langle {\bf{V}}_{\perp} \rangle}{1+\psi}  B_z + \frac{ n_{*}\big(\langle \gamma \rangle{\bf \nabla}_{\perp} \psi -\nabla_{\perp}(|\hat{\bf{A}}_{\perp}|^2/4)\big)}{(1+\psi)^2}  -
  \frac{n_{*}\langle {\bf{V}}_{\perp} \rangle}{1+\psi}  E_z -
  \frac{n_{*}\langle {\bf{V}}_{\perp} {\bf{V}}_{\perp} \rangle}{1+\psi} \cdot {\bf \nabla}_{\perp} \psi
\end{equation}

After substituting the source term ${\bf{S}}$, which does not contain any time-like $\xi$ derivatives, from Eq.(\ref{eq:FluxCurr1_A}) into Eq.(\ref{eq:EqM_N8_AA}), we find that under the QSA, all fields are calculated at each slice $\xi$ from a 2D ("local") nonlinear equation. The details of solving the equation for an expanded field vector $\Psi = \left( \psi, E_z, B_z, {\bf B}_{\perp}\right)^{\rm T}$ in terms of the electron densities $n_{*}$, fluid velocities $\langle {\bf V}_{\perp} \rangle$, and pressure tensor ${\bf T}$ calculated in the same $(x,y)$ plane as the ${\rm \Psi}$-vector are presented in Sec.~\ref{sec_benchmark}. While equations similar to Eqs.~(\ref{eq:EqM_N8_AA}) - (\ref{eq:Mag_x_A}) were presented earlier~\cite{QUICK_2013,HiPACE_2014}, no explicit form of $\frac{\partial}{\partial\xi}{\bf{j}}_{\perp}$ has been presented. We further note that while the $E_z$ component of the expanded field vector does not explicitly enter into the equations of motion of the quasi-statically treated plasma electrons, it does enter the equations of motion of the driver bunch as explained below.

\section{Description of different plasma wake drivers and their interactions}\label{sec_laserfields}
Several types of drivers are implemented in WAND-PIC: laser pulses, charged beams, or both. For a laser driver with the carrier frequency $\omega_0 = k_0c$ and a vector potential $\tilde{\bm{A}}_{\perp} =\hat{\bm{A}}_{\perp}\exp[-ik_0\xi]$, we solve the following paraxial equation for the complex-valued envelope $\hat{\bm{A}}_{\perp}$:
\begin{eqnarray}
\Big(ik_0\frac{2\partial}{\partial t}-\frac{2\partial^2}{\partial t\partial \xi}+\nabla_{\perp}^2\Big)\hat{\bm{A}}_{\perp}=k_p^2\chi\hat{\bm{A}}_{\perp},
\label{eq:laser_Enve}
\end{eqnarray}
where the effective plasma susceptibility averaged over the laser period, $\chi = \big< n_e/\gamma\big>_{2\pi/\omega_0} = n_*/(1+\psi)$, is a local-averaged quantity~\cite{QS_Pulse}.

For the charged bunch drivers, several types of charged macro-particles are enabled in WAND-PIC, including electrons and a variety of ions. Driver particles are different from plasma trajectories in two respects: (i) driver particles are not subject to quasi-static approximation, i.e. their ($x(t),y(t),z(t)$) trajectories are calculated; (ii) in those cases where both the charged beam and the laser pulse drivers are present (e.g., in the context of LEPA~\cite{My_LEPA_2020}), we go beyond the ponderomotive (frequency-averaged) approximation and include the full high-frequency laser fields ($\tilde{\bm{E}}^{L},\tilde{\bm{B}}^{L}$) to advance the driver beam particles. As an example, consider a tightly-focused laser pulse polarized primarily in $x-$direction. The vector potential of such a laser pulse has two components: $\tilde{A}_{x}$ and $\tilde{A}_{z}$ satisfying $|\hat{A}_{x}|>>|\hat{A}_{z}|$. From the vector potential, we obtain the following electric and magnetic field components that are retained in the code:

\begin{eqnarray}
&&\tilde{E}^{L}_x=-\frac{\partial \tilde{A}_{x}}{\partial t}=-\Big(\Big(\frac{\partial \hat{A}_{x}}{\partial t}+\frac{\partial \hat{A}_{x}}{\partial \xi}\Big)-ik_0\hat{A}_{x}\Big)\exp(-ik_0\xi),\\
&& \tilde{B}^{L}_y=\frac{\partial \tilde{A}_{x}}{\partial z}-\frac{\partial \tilde{A}_{z}}{\partial x}\approx\Big(-\frac{\partial \hat{A}_{x}}{\partial \xi}+ik_0\hat{A}_{x}\Big)\exp(-ik_0\xi),\\
&& \tilde{E}^{L}_z\approx\Big(-\frac{\partial \hat{A}_{z}}{\partial \xi}+ik_0\hat{A}_{z}\Big)\exp(-ik_0\xi)=-\frac{\partial \hat{A}_{x}}{\partial x}\exp(-ik_0\xi),\\
&& \tilde{B}^{L}_z=-\frac{\partial \hat{A}_{x}}{\partial y}\exp(-ik_0\xi).
\label{eq:E_B_Fields}
\end{eqnarray}

In deriving the above equations for the laser components we have used the Coulomb gauge $\nabla\cdot\tilde{\bm{A}}=0$ and assumed that the laser spot size is larger than $k_0^{-1}$. Such an approximation enables us to drop the negligibly-small $\tilde{E}^{L}_y$ and $\tilde{B}^{L}_x$ while retaining the small but finite $\tilde{E}^{L}_z$ and $\tilde{B}^{L}_z$ laser field components.  Therefore, for a linear polarized (in $x$-direction) laser pulse, two electric and two magnetic components are determined from one dominant envelope $\hat{A}_x$. If an orthogonal polarization component $\hat{A}_y$ exists, we simply need to solve an additional envelope equation for that component. For every particle of the driver bunch, we calculate and interpolate the four (for a circularly polarized laser: six) laser field components and six wakefield components onto its location. We then use a Boris-like pusher to advance the driver particles while using the sub-cycling method to ensure that the high-frequency fields are properly resolved in time. Thus, when we have both laser driver and beam driver overlapped inside the bubble, i.e., in the case of the DLA of a witness bunch~\cite{Xi_prl,zhang_ppcf,shaw_ppcf,shaw_ppcf2,zhang_ppcf_2,Kh_2018,My_DLA_2019} or a LEPA scheme~\cite{My_LEPA_2020}, the interaction between the laser and the driver (or witness) particles can be accurately modeled by WAND-PIC.

For completeness, we list the full equations of motion for a $j$'th particle of the driver bunch under the influence of the laser and wake fields:
\begin{eqnarray}{}
&&\frac{d}{dt} {\bf R}_j (t) = \frac{{\bf P}_j(t)}{\gamma_j} \label{eq:EoM_beam_1}, \\
&&\frac{d}{dt} {\bf P}_j(t) = \frac{Q_j}{M_j}\left( \tilde{\bm{E}}^{L}({\bf R}_{j}(t),t)+\bm{E}^{W}({\bf R}_{j}(t),t)\right) + \frac{Q_j}{M_j}\frac{{\bf P}_j(t)}{\gamma_j} \times\left( \tilde{\bm{B}}^{L}({\bf R}_{j}(t),t)+\bm{B}^{W}({\bf R}_{j}(t),t)\right),\label{eq:EoM_beam_2}
\end{eqnarray}
where $\gamma_j=\sqrt{1+{\bf P}^2_j(t)}$, $Q_j$ and $M_j$ are normalized charge and mass of the $j$'th driver particle, respectively. Wakefields $\bm{E}^{W}={\bf{e}}_zE_z+\bf{E}_\perp$ and $\bm{B}^{W}={\bf{e}}_zB_z+\bf{B}_\perp$. The $E_z$, $B_z$, and ${\bf B}_{\perp}$ components of the wakefield are contained in the expanded wakefield vector $\Psi$, and the transverse electric field is calculated as $\bf{E}_\perp=-\nabla_\perp\psi +\bf{e}_z\times\bf{B}_\perp$.

\section{Comparison with full 3D PIC simulations}\label{sec_compare}
In this section, we compare the performance of WAND-PIC -- in terms of accuracy and computational efficiency -- with that of the full PIC code VLPL. Because WAND-PIC is based on several key assumptions and approximations, the following questions will be answered. First, we will test the efficacy of iterative step size refinement by modeling the wakefield in the back of a plasma bubble driven by a luminal ($v_b = c$) non-evolving high-charge electron bunch (see Sec.~\ref{sec_beam_driver}). Then, we will validate the efficacy of using the sub-cycling algorithm to accurately model DLA effects in LEPA~\cite{My_LEPA_2020}, where an electron bunch and a laser pulse driver co-propagate in the plasma. 
\subsection{Plasma bubble driven by a beam driver with large charge}\label{sec_beam_driver}
Below we assess the performance of WAND-PIC in simulating the wakefield driven by an electron beam driver with a large charge $q$, defined according to $Q \equiv k_p^3q/(4\pi en_0) \gg 1$ ~\cite{stupakov_2016,My_Driver_2017}. It has been well-established~\cite{Lu-Beam-Theory,My_Driver_2017} that the back of the bubble contains highly-relativistic electron trajectories and a steeply-profiled wakefield. The maximum momenta of an electron with a trajectory along the bubble's edge have been estimated~\cite{Injection_Theory} as $p_r \approx Q$ and $p_z \approx p_x^2/2=Q^2/2$. This analytic estimate illustrates the challenge of simulating such a bubble with a quasi-static code: the longitudinal "velocity" $v_\xi=1-p_z/\gamma\approx 2/Q^2$ of such a particle in the reference frame co-moving with the bubble is much smaller than its transverse velocity $v_x\approx2/Q$. Therefore, the longitudinal step size $\delta\xi$ should be smaller than the transverse one $\delta x$, thus suggesting that $\delta\xi \leq \delta x/Q$ can be chosen to accurately capture the steepness of the back regions of the bubble. To make the calculation more efficient, we calculate the required $\delta\xi$ adaptively at every $\xi$ step according to the maximum transverse velocity of all trajectories: $\delta\xi\propto 1/max(|\bf{V}_{\perp}|)$, in this way, we make sure that finer meshes are deployed only near the back of the bubble.

\begin{figure}[htp!]
\centering
\includegraphics[width=0.8\columnwidth]{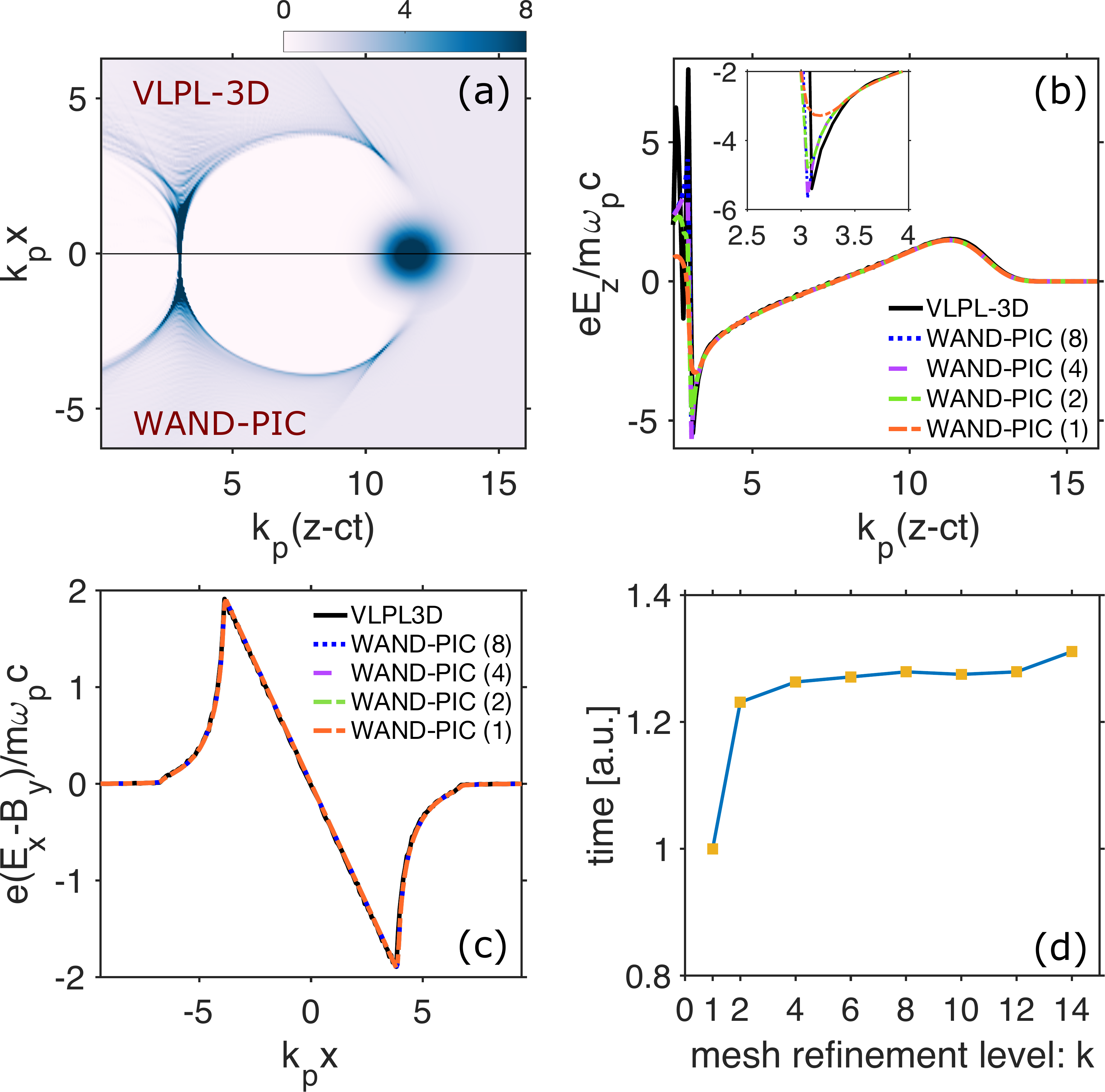}\\
\caption{Comparison of the simulation results from VLPL-3D and WAND-PIC. (a) Plasma bubbles from VLPL-3D (upper half) and WAND-PIC (lower half). (b) Longitudinal on-axis wakefield $E_z(\xi,{\bf r}=0)$ from VLPL-3D and WAND-PIC ($k$). Maximum step size refinement level $k$: step size refinement proceeds until the adaptive step size $d\xi$ is reduced by $k$ from the original step size. (c) Transverse focusing wakefields $F_{\perp x}(x) = E_x - B_y$ at the center of the bubble in the $x-z$ plane. (d) Normalized runtime of WAND-PIC ($k$) for different levels of step size refinement $k$. }
\label{Fig2}
\end{figure}

In the following simulation, we choose a uniform plasma density $n_0=6.5\times10^{17}cm^{-3}$ and a $10$GeV electron beam with Gaussian charge density distribution: $n_b=10n_0 e^{-r^2/\sigma_r^2 - z^2/\sigma_z^2}$, where $\sigma_r = 6.6\mu m$ and $\sigma_z = 8\mu m$. The total $q=2$nC charge of the electron beam corresponds to the normalized charge $Q=5.34$ and the peak current $I=45.5 {\rm kA}$.  Such a beam is not beyond the reach of modern accelerators because electron beams with $q \sim 2 {\rm nC}$ and $I  \sim 15 {\rm kA}$ are already available at the FACET-II facility at SLAC National Accelerator Laboratory~\cite{FACET-II-2020}, and electron beams with currents of $50\mbox{--}150{\rm kA}$ and durations of $3{\rm fs}$ will be available in the near future~\cite{FACET-II-Design}.

We simulate this beam-plasma configuration with WAND-PIC and VLPL-3D, and compare their results in Fig.~\ref{Fig2}. The simulation box sizes in both codes are chosen as $L_x\times L_y\times L_z=4\lambda_p\times4\lambda_p\times2.7\lambda_p$, and the spatial resolution is chosen as $0.01\lambda_p$ in all three dimensions. In Fig.~\ref{Fig2}~(a), the plasma bubbles generated by WAND-PIC and VLPL-3D are compared side-by-side in the $x-z$ plane. The two simulations are clearly in good agreement in terms of the bubble length and radius, as well as the location and steepness of the bubble closure in its back region. Some small differences can be observed: for example, the bubble-bounding electron sheath is narrower in WAND-PIC, and the beginning of the second bubble in WAND-PIC is slightly larger.

One of the most important features of a large bubble is its highly-nonlinear wakefield that must be accurately calculated because the peak accelerating field is essential for accurate estimates of the final energy gain and quality of the witness bunch. Therefore, $\xi$-dependent step size $\delta\xi \equiv \delta\xi (\xi)$ must be used, with much smaller $\delta \xi$ near the back of the bubble than the initial step size $\delta \xi^{(0)}$ at the front of the bubble. To demonstrate the effect of step size refinement, we define the refinement level $k$ as follows: the smallest step size $\delta\xi_{\min}$ of the simulation satisfies $\delta\xi_{\min} \geq  \delta \xi^{(0)}/k$. In Fig.~\ref{Fig2}~(b), we compare the on-axis $E_z$ from VLPL-3D and four different WAND-PIC simulations with different step size refinement levels labeled as WAND-PIC ($k$). For example, WAND-PIC ($1$) means that the step size refinement is turned off, and $\delta \xi$ always equals the initial step size $\delta \xi^{(0)}$.

From Fig.~\ref{Fig2}~(b) and its inset, we can see that WAND-PIC with step size refinement level $k\geq4$ and VLPL-3D produce close results (difference $\approx 5\%$). Lower refinement level ($k=2$) generates $15\%$ smaller peak $E_z$, and WAND-PIC without step-size refinement ($k=1$) generates $40\%$ smaller peak $E_z$. A higher refinement level enables calculating electrons' speeds and the wakefields more accurately, and convergence is eventually reached for increasing $k$: the WAND-PIC with $k=4$ and $k=8$ generate the same results as observed from Fig.~\ref{Fig2}~(b). Empirically, we find that $k=4\sim8$ is sufficient for most of the PWFA simulations. Because trajectories only acquire relativistic speed at the back of the bubble, step size refinements are primarily required in a small region in the back of the bubble, where the electron sheath is very narrow. Therefore, the wakefields everywhere other than at the back of the bubble are accurately solved even without step-size refinement. As shown in Fig.~\ref{Fig2}~(c), the focusing fields $F_{\perp x} = E_x - B_y$ at the bubble center obtained from VLPL-3D and WAND-PIC are near-identical regardless of the step size refinement. For the same reason, the overall runtime of the code does not significantly increase even for $Q \gg 1$ as we increase the level of step size refinement. The runtime of WAND-PIC with different $k$ plotted in Fig.~\ref{Fig2}~(d) shows that most of the runtime increase (by $\approx 20\%$) takes place between $k=1$ and $k=2$. As $k$ increases from $k=2$ to $14$, the runtime barely increases.

This example shows that the adaptive step size refinement in WAND-PIC is a useful and necessary technique for simulating large plasma bubbles in PWFAs. The effectiveness of using adaptive step size refinement is particularly high when all wakefield equations are local in $\xi$, i.e. do not contain any sources containing $\partial_\xi$: straightforward integration of plasma electron trajectories in the $\xi$-direction is always stable as long as the step size is appropriately refined. In contrast, non-local quasi-static codes utilizing the predictor-corrector approach often exhibit unstable performance for large bubbles driven by the $Q\gg1$ charges.

In fact, it is not uncommon to encounter physical situations where predictor-corrector codes fail to converge due to numerical instability at the back of the bubble even for very small step sizes $d\xi$ and a large number of predictor-corrector iterations. One common approach is to use a fairly large $d\xi$ and the $1^{st}$ order predictor-corrector scheme to work around this issue. However, such an approach underestimated electron speeds at the back of the bubble. Another work-around frequently used in combination with the predictor-corrector approach is to add a speed limitation on plasma electrons, or to avoid simulating the back of the bubble altogether by truncating the size of the computational domain. Unfortunately, none of these methods produce sufficiently accurate wakefields at the back of the bubble as long as the normalized drive charge $Q$ is large. On the other hand, driver charges corresponding to $Q \sim 100$ have been successfully simulated with WAND-PIC~\cite{My_Driver_2017}.

\subsection{Modeling direct laser acceleration of electrons with WAND-PIC}\label{sec_DLA}
Next, we used WAND-PIC to simulate the direct laser acceleration (DLA) of electrons in the bubble regime and compare the results with the full PIC code VLPL-3D. In the context of LWFA, the electrons inside a plasma channel or plasma bubble can gain energy directly from the wake-generating (pump) laser pulse ~\cite{shaw_ppcf,shaw_ppcf2} or an additional trailing (DLA) pulse~\cite{Xi_prl,zhang_ppcf,zhang_ppcf_2} added to the location of the electron bunch pre-injected into the back of the plasma bubble produced by the pump pulse. The DLA mechanism can happen in a long plasma channel where longitudinal wakefield is zero~\cite{Khudik_2016}, or in the accelerating portion of the bubble~\cite{shaw_ppcf,Xi_prl}, or even in the decelerating portion of the bubble~\cite{Kh_2018}, as long as the Doppler-shifted frequency $\omega_D = \omega_L(1-v_z/v_{\rm ph})$ of the laser field matches the electrons' betatron frequency $\omega_{\beta}=\omega_p/\sqrt{2\gamma}$ in the channel/bubble. Here $\omega_L$ and $v_{\rm ph}$ are the laser frequency and phase velocities, respectively, and $v_z$ is the longitudinal electron velocity. The actual description is complicated by the fact that the $\langle \omega_D \rangle = \langle \omega_{\beta} \rangle$ relationship is only satisfied on average~\cite{shaw_ppcf,Khudik_2016} because of the rapid nonlinear variation of $v_z$ during one betatron period $T_{\beta} = 2\pi /\omega_{\beta}$. The result of such relativistic nonlinearity of the laser-particle interaction is an irregular (stochastic) motion of the accelerated electrons~\cite{Krash_2018}. On the modeling side, the energy $W_{\perp}$ gained by the electrons directly from the laser is highly sensitive to the spatial-temporal resolution of the PIC code used to model the DLA process~\cite{DLA_require_Alex}. This places serious constraints on the length of DLA-based schemes that can be modeled using full PICs, and creates opportunities for reduced-description modeling using quasi-static codes such as WAND-PIC.

\begin{figure}[htp!]
\centering
\includegraphics[width=0.9\columnwidth]{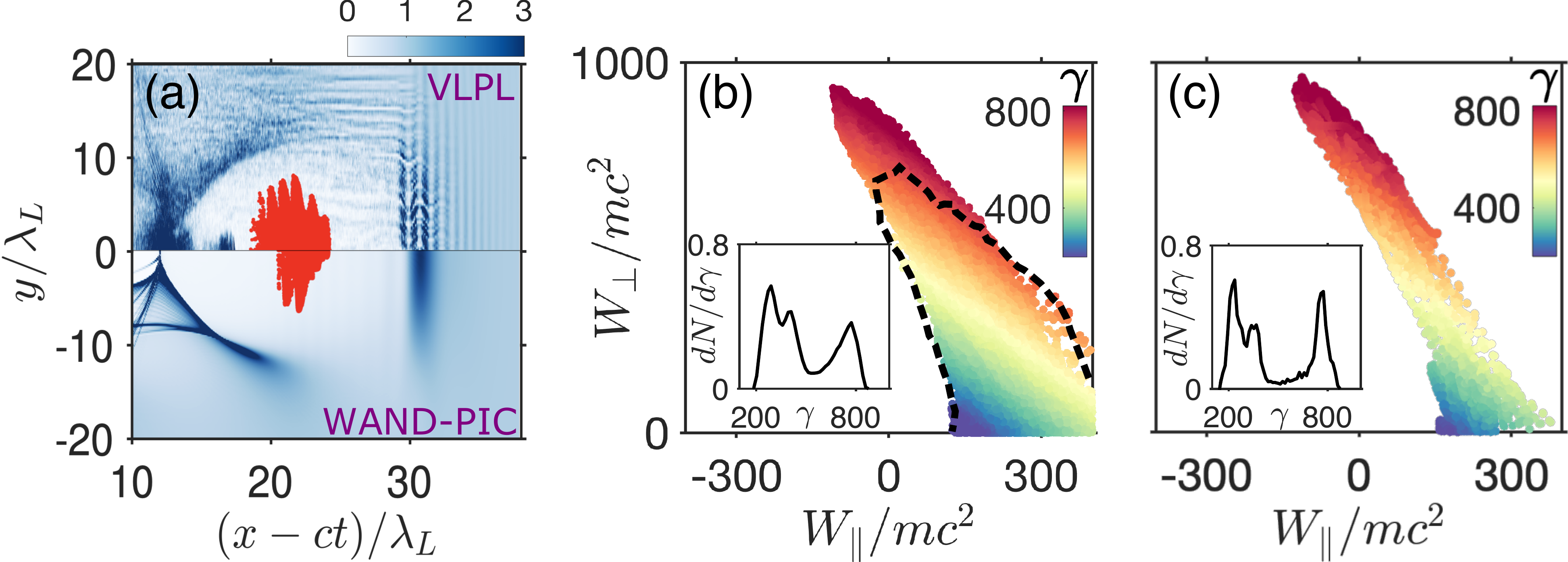}\\
\caption{Direct laser acceleration of an injected electron bunch by a leading (pump) and trailing (DLA) pulses: comparison between VLPL-3D (top of (a) and (b) panels) and WAND-PIC (bottom of (a) and (c) panels). (b) Densities of the plasma electrons and externally injected electrons (red). (b, c) Distribution of the injected electrons in the $(W_{\parallel},W_{\perp})$ space of the work done by the longitudinal ($W_{\parallel}$) and transverse ($W_{\perp}$) electric fields. Color-coding: by relativistic factor $\gamma$, insets: electron energy spectra. Dashed black curve in (b) encloses the $(W_{\parallel},W_{\perp})$ space from the lower-resolution VLPL-3D simulation. Laser parameters: peak powers $P_{\rm pump}=P_{\rm DLA} = 21 \rm TW$, wavelengths $\lambda_{\rm pump} = \lambda_{\rm DLA} = 0.8\mu$m, durations $\tau_{\rm pump}=16.6 \rm fs$ and $\tau_{\rm DLA}=9.4$fs, spot sizes $w_{\rm pump}=8.7\mu$m and $w_{\rm DLA}=5.5\mu$m. Plasma parameters: density $n_0 =7.7 \times 10^{18} \rm cm^{-3}$and length $z=1.15$mm. Grid/step size for high (low) VLPL-3D resolutions: $\delta x=\delta y=\lambda_L/5,~\delta z\approx c\delta t=\lambda_L/100$ ($\delta x=\delta y=\lambda_L/5,~\delta z\approx c\delta t=\lambda_L/50$).}\label{Fig3}
\end{figure}

The accuracy of modeling DLA in the bubble regime with WAND-PIC was tested using the following two-pulse setup~\cite{Xi_prl}. A leading pump and a trailing DLA pulses polarized in $x$ direction and separated by the time delay of $\Delta \tau = 33$fs are launched into a tenuous plasma, see Fig.~\ref{Fig3} caption for laser and plasma parameters. The two pulses co-propagate through the plasma over a dephasing distance $z=L_{\rm deph} \approx 1.15 \rm mm$ with a short electron bunch pre-injected into the center of the DLA pulse. The initial momentum of the bunch $p_z = 15mc$. The bunch duration $\tau_b = 4 \rm fs$ and transverse size $w_{\rm bunch}= 3\mu \rm m$ are chosen, and its total electric charge is assumed to be negligible. The energy gains of the accelerated electrons from the longitudinal ($W_{\parallel}$) and transverse ($W_{\perp}$) electric fields are calculated and plotted in Fig.~\ref{Fig3}(b) for VLPL-3D and in Fig.~\ref{Fig3}(c) for WAND-PIC simulations. Two separate VLPL simulations were carried out: the high-resolution ($\delta x=\delta y=\lambda_L/5,~\delta z\approx c\delta t=\lambda_L/100$) and the low-resolution ($\delta x=\delta y=\lambda_L/5,~\delta z\approx c\delta t=\lambda_L/50$). For the WAND-PIC simulations, the mesh and step sizes are as follows: $\delta x=\delta y=\lambda_L/6.28,~\delta \xi=\lambda_L/16.7,~c\delta t=\lambda_L/2$, and the sub-cycling number $N_{sub}=50$ for the witness bunch. We note that $\delta \xi$ and $\delta t$ in WAND-PIC are significantly larger than $\delta z$ and $\delta t$ in VLPL-3D.

This particular set of laser-plasma parameters is particularly interesting because the tightly focused/guided DLA pulse in the back of the bubble has a non-vanishing longitudinal electric field component:  $|E^{L}_{\parallel}|\propto x|E^{L}_{\perp}|/(k_0R^2)$~\cite{pukhov2002_DLA,My_DLA_2019}, where $x$ is the transverse coordinate. For the bubble radius $R\sim k_p^{-1}$ and a relativistic laser pulse $a_L \equiv e|E^{L}_{\perp}|/mc\omega_0 > 1$, the electrons undulating with betatron amplitude $\sim R$ can experience comparable longitudinal electric fields from the wake and the laser pulse. An accurate simulation of DLA in such a regime contributes to a better understanding of the energy transfer to the accelerated electrons from the wake and laser fields. We note that separating the two contributions to electron energy is much more challenging for the full PIC codes because all electric fields, including those of the wake and the laser, are combined. On the other hand, WAND-PIC separates the two.

Figure\ref{Fig3}~(a) presents a side-by-side comparison of the plasma bubbles and pre-injected electron bunches simulated by the high-resolution VLPL-3D and WAND-PIC code in the upper- and lower halves of the figure, respectively. We observe that the bubble sizes are very similar in both simulations. However, while a small quantity of self-injected electrons can be observed at the back of the bubble simulated with VLPL-3D, self-injection falls outside of the QSA and is not allowed in WAND-PIC. One consequence of this is that the bubble boundary is more clearly defined in WAND-PIC simulations. We further observe good agreement between VLPL-3D and WAND-PIC simulations of the externally-injected electron bunches in that they have similar transverse sizes (approximately doubled from their original size under the action of the DLA), and have both advanced to the same positions inside the bubble. The electron bunch in WAND-PIC is also found to be less stretched in the longitudinal direction.

To evaluate the energetics of the DLA process in both codes, the energy transfer phase space $(W_{\parallel}, W_{\perp})$ for the externally-injected electrons is plotted in Figs.~\ref{Fig3}~(b,c) for the simulations by high-resolution VLPL-3D and WAND-PIC, respectively. The energy transfers $W_{\parallel}$, and $W_{\perp}$ are defined as follows:
\begin{eqnarray}
  &&W_{\parallel} \equiv A_{z}^{\rm L} + A^{\rm W} =   -e \int \left( E_z^{\rm L} + E_z^{\rm W} \right)~v_z dt \label{eq:W_parallel},\\
  &&W_{\perp} \equiv A_{\perp}^{\rm L} = -e \int \mathbf{E}_{\perp} \cdot \mathbf{v}_{\perp} dt \approx -e \int E_x^{\rm L}~v_{x} dt  ,  \label{eq:W_perp}
\end{eqnarray}
where $E_z^{\rm L}$ and $E_x^{\rm L}$ are the longitudinal and traverse laser electric fields, respectively, and $E_z^{\rm W}$ is the longitudinal wakefield. Similarly, $A^{\rm W}$ is the work done by the wakefield, and $A_{\perp,z}^{\rm L}$ is the work done by the transverse (longitudinal) components of the laser field. Note that, while the work $A^{\rm W}$ done by the wake and the work $A^{\rm L} = A_{\perp} + A_{z}$ done by the laser are of considerable theoretical interest, only the $W_{\parallel}$ and $W_{\perp}$ quantities can be cross-checked for the two codes because $A^{\rm W}$ and $A^{\rm L}$ cannot be separately calculated by the VLPL-3D.

As shown in Fig.~\ref{Fig3}~(b) and (c), the results from VLPL-PIC and WAND-PIC are in excellent agreement with each other. Not only the absolute gains in longitudinal and transverse directions are close in the two codes, but also the phase space distributions are in good agreement. This indicates that the WAND-PIC is accurately modeling the field components of the laser, as well as the interaction between the laser and electrons through the sub-cycling method. The insets in Fig.~\ref{Fig3}~(b) and (c) show that the energy spectra of all electrons obtained from VLPL-3D and WAND-PIC are in good agreement in terms of the energy range and their spectral shapes (e.g., three-peak features from both codes). Note that the black-dashed contour in Fig.~\ref{Fig3}~(b) encloses the phase space $(W_{\parallel}, W_{\perp})$ obtained from the low-resolution VLPL-3D. In that simulation, the energy gain from the transverse electric field of the laser is underestimated by $30\%$; this would influence the final energy distribution of the electrons, as well as the radiation output associated with betatron oscillation. Therefore, Fig.~\ref{Fig3}~(b) informs us that, in order to accurately capture the complex interactions between electrons and laser fields, a full PIC code needs a longitudinal/time resolution around $\delta z \sim \lambda_L/100$. However, the WAND-PIC is able to simulate the bubble evolution and DLA mechanism separately by using coarser resolution for the laser and sub-cycling for the electrons. For the specific example shown in Fig.~\ref{Fig3}, the WAND-PIC uses two orders of magnitude fewer computer resources (as measured in core-hours) than the high-resolution VLPL-3D.

\section{Example Simulation: Description of Phase-dependent non-axisymmetric bubble undulations with WAND-PIC}\label{CEP_PUB}
In this section, we demonstrate the ability of WAND-PIC to capture subtle three-dimensional phase-dependent effects from the laser pulse. While phase-related phenomena are discussed in the context of relatively dense plasmas (e.g., $\omega_L/\omega_p \sim 10$), for example, the carrier envelope phase (CEP) effect where the difference between group velocity and phase velocity of laser is important~\cite{CEP_observable,Zhengyan,Salehi,kost_cep,Jihoon}, there could be other accelerator-relevant circumstances where laser phase is relevant. For example, an ultra-short bunch with duration $\tau_b < \lambda_L$ externally injected at the laser intensity peak can undergo DLA-induced transverse undulation. Because such undulation is phase-dependent, it produces a phase-controlled asymmetric deflection of plasma electrons~\cite{kost_cep}, thereby inducing phase-dependent undulations of a bubble (PUB)~\cite{Jihoon} along the direction of laser polarization. Numerical description of such PUBs within the QSA framework requires a fully-3D description of the plasma wake and accurate modeling of the sub-luminal driver: a laser pulse propagating with $v_g < c$. Here we demonstrate that such sub-luminal non-axisymmetric plasma wakes are accurately described by WAND-PIC.

\begin{figure}[htp!]
\centering
\includegraphics[width=0.8\columnwidth]{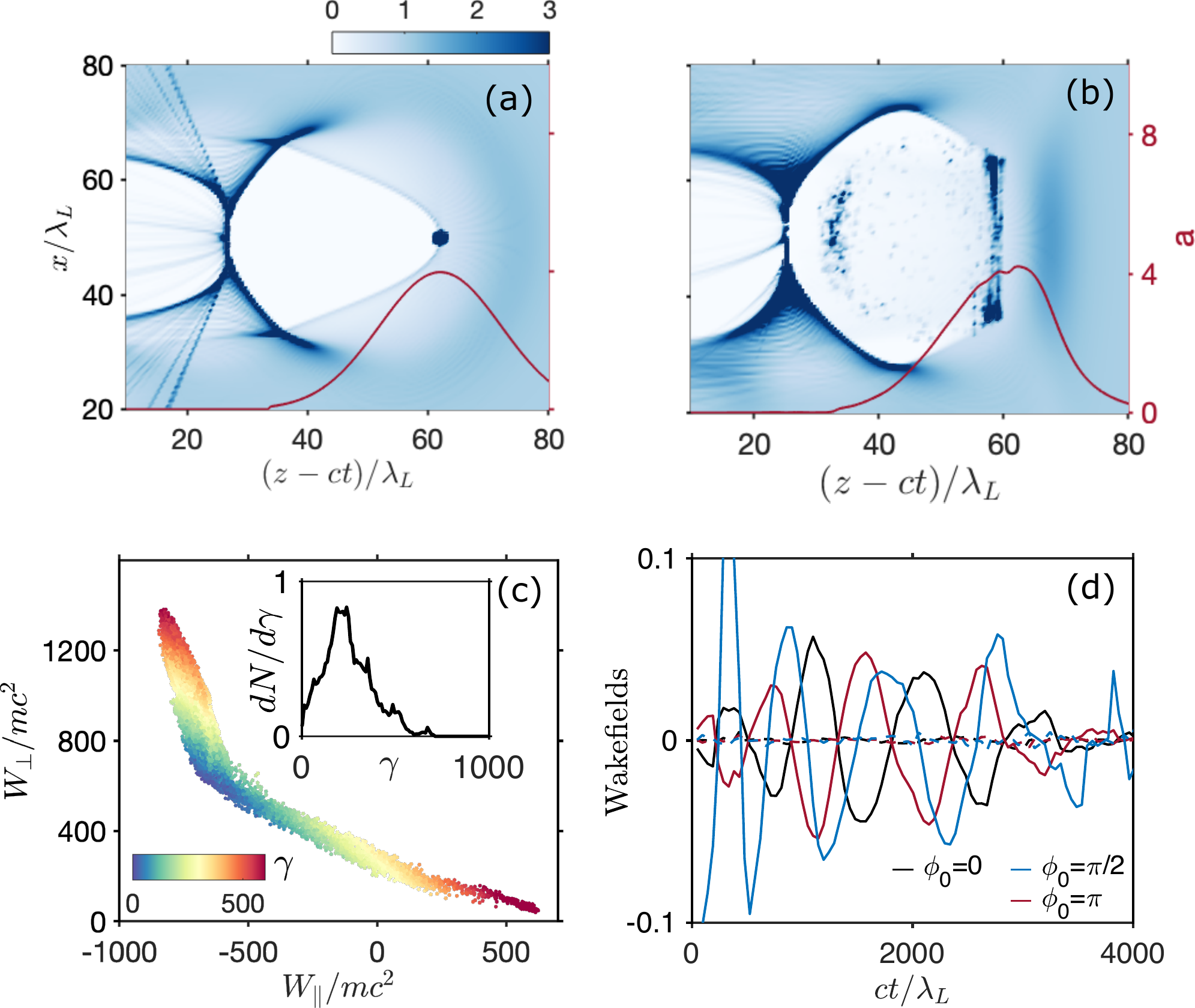}\\
\caption{3D simulation done by WAND-PIC. (a) Plasma bubble and electron bunch at propagation distance $z=0$mm, the red curve represents the laser envelope. (b) Same as (a), but at propagation distance $z=25.2$mm. (c) Bunch electrons' phase space distribution ($W_\parallel,W_\perp$) at distance $z=25.2$mm, the inset shows the energy spectrum at same moment. (d) Transverse wakefields at the back of bubble: $(z-ct)/\lambda_L=30$. The solid lines show the transverse wakefields in laser polarization direction $F_x=E_x-B_y$, and the dashed lines show $F_y=E_y+B_z$. Different colors stand for different initial laser phase, black: $\phi_0=0$, blue: $\phi_0=\pi/2$, and red: $\phi_0=\pi$.  Simulation parameters: $\delta x=\delta y=\delta z=c\delta t=0.1k_p^{-1}$, sub-cycling number $N_{sub}= 50$.}
\label{Fig4}
\end{figure}

The simulation setup is shown in Fig.~\ref{Fig4}~(a), a linear polarized ${\rm CO}_2$ laser pulse with wavelength $\lambda_L=9.2 {\rm mm}$ and peak $a_0=4.0$ is propagating in plasma with density $n_0=1.46\times10^{16}{\rm cm}^{-3}$. The laser's duration $\tau_{L}=550{\rm fs}$, spotsize $w=132 \mu m$, and power $P=142$TW. For such a laser duration: $\sim18\lambda_L$  and plasma density: $\omega_p=\omega_0/30$, the self-steepening of the laser pulse is not strong enough to produce a sharp laser front, therefore, it's unlikely that this pulse will introduce measurable periodic bubble undulation\cite{Jihoon,Jihoon2} throughout its propagation. Therefore, a short electron bunch with initial energy $15{\rm MeV}$ is placed at the peak of the laser envelope to induce a phase-dependent undulation. The charge of bunch $q_b=2.3$nC which corresponds to a normalized charge $Q=0.92$, and the bunch duration is $\tau_b=15{\rm fs}$. Although the bunch electrons are placed at the decelerating phase of the bubble, they gain energy from the laser through the DLA mechanism that overcomes the deceleration by the plasma wakefield~\cite{Kh_2018,My_LEPA_2020}. Since the bunch duration is about half of the laser wavelength, bunch electrons would have some degree of synergy in the DLA process, and the oscillation of the bunch, with a frequency roughly equal to the betatron frequency, would shake the bubble periodically. 

The DLA is captured accurately by our particle pusher with the sub-cycling technique. At propagation distance $z=25.2{\rm mm}$, bunch electrons gained significant energy from the laser as shown in Fig.~\ref{Fig4}~(c). The average gain from the laser pulse $\langle W_\perp\rangle$ is about 400MeV, and the average loss to the longitudinal field $\langle W_\parallel\rangle$ (laser component and wakefield combined) is 300MeV. The DLA also results in the broadening of the bunch in the laser polarization direction (x) as shown in the Fig.~\ref{Fig4}~(b). The oscillation triggers the undulation of the bubble in the x-direction and this undulation is phase-dependent. When measuring the transverse wakefields $F_x=E_x-B_y$ and $F_y=E_y+B_z$ at the back of the bubble, we see from Fig.~\ref{Fig4}~(d), that $F_x$ (solid lines) show periodic oscillation while the $F_y$ (dashed lines ) are negligible. The amplitudes of $F_x$ are about $0.72{\rm GeV/m}$. Different colors in Fig.~\ref{Fig4}~(d) correspond to the different initial phases of the laser: $\phi_0$. The oscillation of the $F_x$ also shows a strong correlation with the laser phase. Together with the polarized undulation direction, these results demonstrate that this effect is indeed DLA-induced and phase-dependent.

This particular setup shows the importance of the full 3D geometry and the modeling of DLA in WAND-PIC. In the presence of both laser pulse and electron bunch, the phase-dependent effect is captured through the correct modeling of sophisticated electron-laser interactions, and the visualization of this effect is facilitated by the non-axisymmetric plasma flows in the full 3D geometry. 

\section{Algorithm, efficiency and scaling}\label{sec_benchmark}

In this section, we discuss the algorithm design and the benchmarking of the WAND-PIC. First, the global routine of WAND-PIC is discussed. Then we briefly present the implementation of the non-uniform transverse grids in our MG solver and show one sample application of this feature. At last, we discuss the efficiencies and scalings of WAND-PIC on the distributed parallel computing system.
\subsection{WAND-PIC's Global Algorithm}

The algorithm design and the code structure of WAND-PIC is greatly simplified by the local in-$\xi$ method of calculating the wakefield components as described by Eqs.(\ref{eq:EqM_N8_AA},\ref{eq:EqM_N8_A},\ref{eq:Mag_z_A},\ref{eq:Main_Eq_A},\ref{eq:FluxCurr1_A}) which are solved without using predictor-corrector schemes. Therefore, the entire algorithm consists of two main loops: (i) the time loop in $t$ for the driver(s), and (ii) the slicing loop in $\xi$ for the plasma electrons and wakefields, as shown in pseudocode~(\ref{algo1}): 

\begin{algorithm}
	\caption{WAND-PIC's Global Algorithm} 
	\begin{algorithmic}[1]
	\State Initialization()
		\For {each time step $t=t_i$}
			\For {each $\xi$ step $\xi=\xi_j$}
				\State Collect{\_}Source()
				\State Solve{\_}Wakefield()   \Comment{depending on the pushing method, particle pusher and field solver can be intertwined.}
				\State Push{\_}Trajectory()
				\State Step{\_}Size{\_}Refinement()
			\EndFor
			\State Push{\_}Driver()
		\EndFor
\end{algorithmic} 
\ 
\ 
\begin{algorithmic}[1]
	\Function{{\rm Step{\_}Size{\_}Refinement()}}{}
    	\State $V_{\max} = \max(|{\bm V}_\perp|~of~all~trajectories)$
    	\State $V_{0}=\delta x/\delta\xi^{(0)}$
    	\If{$V_{\max}>V_0$}
    		\State	$\delta\xi  = \delta\xi^{(0)}*V_0/V_{\max}$   \Comment{this makes sure trajectories never across more than one transverse grid.}
    	\Else
        	\State	$\delta\xi = \delta\xi^{(0)}$
    	\EndIf
    \EndFunction
	\end{algorithmic} 
	\label{algo1}
\end{algorithm}

Since the time-loop is relatively straightforward, we now focus on the slice-loop in $\xi$ . At every time step $t=t_i$, the drivers are assumed "frozen" in the moving frame, and plasma flows -- macroparticles and wakefields -- are advanced in the positive $\xi$-direction. The routine at one $\xi$  slice consists of main three steps: (i) collecting source, (ii) solving fields, and (iii) advancing particles. At every $\xi$ slice, the source terms -- currents $\langle {\bf V}_\perp\rangle$, number densities $n_*$, and pressure tensor components ${\bf T}$ of the plasma electrons --  are deposited onto the transverse grids, and wakefields are solved. Then wakefields are interpolated to macroparticles' positions, and macroparticles are pushed to the next slice according to Eqs.~(\ref{eq:EoM_112_A},\ref{eq:EoM_111}). A Boris-like pusher~\cite{Boris_pusher} is used to advance macroparticles (1st, 2nd, and 4th-order Runge-Kutta are also available). 

The fields are obtained by solving a 2D nonlinear elliptic equation for the earlier introduced expanded wakefield vector $\Psi$, The equation takes the following form:
\begin{equation}\label{eq:Psi_Matrix}
  {\bf \nabla}_{\perp}^2 \Psi + {\bf M}_1 \cdot \Psi + {\bf M}_{2x} \cdot {\partial}_x \Psi+{\bf M}_{2y} \cdot {\partial}_y \Psi = {\bf J}\left(\Psi; n_{*},\langle {\bf V}_{\perp} \rangle,{\bf T} \right)
\end{equation}
where ${\bf M}_1$ and ${\bf M}_{2x,2y}$ are $5 \times 5$ matrices that depends on $\Psi$ and $\langle {\bf V}_{\perp} \rangle$. ${\bf J}$ is an expanded current source vector defined by the electron densities $n_{*}$, fluid velocities $\langle {\bf V}_{\perp} \rangle$, and pressure tensor ${\bf T}$, all calculated in the same $(x,y)$ plane as the ${\rm \Psi}$-vector. The expressions for $\Psi$, ${\bf M}_{1,2x,2y}({\bf r},\xi)$ and ${\bf J}({\bf r},\xi)$ are as follows:
\begin{equation}\label{eq:matrix}
{\Psi}=\begin{bmatrix}
\psi\\
E_z\\
B_z\\
B_x\\
B_y
\end{bmatrix},
\ \ \ {\bf J} = 
\begin{bmatrix}
n_*-1\\
-\nabla_{\perp}\cdot{\bf{j}}_{\perp}\\
{\bf{e}}_z\cdot[\nabla_{\perp}\times {\bf{j}}_{\perp}]\\
\partial_y\left(j_z-n_*\langle T_{yy}\rangle\right)-\partial_x(n_*\langle T_{xy}\rangle)-n_*\frac{\partial_y|\hat{\bf{A}}_{\perp}|^2}{4(1+\psi)^2}\\
\partial_x\left(n_*\langle T_{xx}\rangle-j_z\right)+\partial_y(n_*\langle T_{xy}\rangle)+n_*\frac{\partial_x|\hat{\bf{A}}_{\perp}|^2}{4(1+\psi)^2}
\end{bmatrix},
\ \ \ {\bf M_1} =n_*
\begin{bmatrix}
0&0&0&0&0\\
0&0&0&0&0\\
0&0&0&0&0\\
0&\frac{ \langle V_y\rangle}{1+\psi}&-\frac{\langle V_x\rangle}{1+\psi}&0&-\frac{1}{1+\psi}\\
0&-\frac{\langle V_x\rangle}{1+\psi}&-\frac{\langle V_y\rangle}{1+\psi}&0&-\frac{1}{1+\psi}\
\end{bmatrix},
\end{equation}
\begin{equation}
\ \ \ {\bf M_{2x}}=n_*
\begin{bmatrix}
0&0&0&0&0\\
0&0&0&0&0\\
0&0&0&0&0\\
\frac{\langle T_{xy} \rangle}{1+\psi}&0&0&0&0\\
-\frac{\langle T_{xx} \rangle}{1+\psi}+\frac{\langle\gamma\rangle}{(1+\psi)^2}&0&0&0&0\
\end{bmatrix},
\ \ \ {\bf M_{2y}} =n_*
\begin{bmatrix}
0&0&0&0&0\\
0&0&0&0&0\\
0&0&0&0&0\\
\frac{\langle T_{yy} \rangle}{1+\psi}-\frac{\langle\gamma\rangle}{(1+\psi)^2}&0&0&0&0\\
-\frac{\langle T_{xy} \rangle}{1+\psi}&0&0&0&0\
\end{bmatrix}.
\end{equation}
For simplicity, the parametric dependences of $\Psi({\bf r},\xi)$ and ${\bf J}({\bf r},\xi)$ on the time-step label $t_i$ is suppressed.  When solving Eq.(\ref{eq:Psi_Matrix}), $\psi$, $E_z$, and $B_z$ are solved first, then their values enter ${\bf M}_1$ and ${\bf M}_{2x,2y}$ matrices to solve the ${\bf B}_\perp$.

While the inhomogeneous Eq.(\ref{eq:Psi_Matrix}) can be solved using a variety of numerical solvers (including the commonly used FFT solver implemented in QuickPIC and HiPACE), WAND-PIC uses an MG solver. The advantages of the multigrid (MG) solvers over the FFT solvers are as follows. First, an MG solver naturally fits into the 2D square partitioning of the $(x,y)$ computational domain, and the communications between different sub-domains are all local: each sub-domain only communicates with the four neighboring ones as shown in Fig.~\ref{Fig1}. This lends MG solvers to efficient parallelization. On the contrary, a 2D FFT solver requires global communications between sub-domains, resulting in considerable efficiency loss under parallelization. Second, MG solvers have potentially better scaling of the computational time (also known as time complexity) with the problem size $n$: $O(n)$ for the MG vs $O(nlog(n))$ for the FFT method. Third, because MG is an inherently iterative method, a good initial guess of the solution can be applied to accelerate the convergence. For example, in WAND-PIC we use the fields from the previous $\xi$ step (or from the previous time step when solving Eq.(\ref{eq:laser_Enve}) for the laser pulse envelope) as the initial guess. We found that an MG solver with an appropriate initial guess is two times faster compared with a trivial ($\Psi = 0$) initial guess.

Once the wakefields are solved and macroparticles are advanced, we run the step size refinement to determine the appropriate step size $d\xi$ used in the next $\xi$ slice. The step size refinement function is explained in the pseudocode~(\ref{algo1}).  After the completion of the $\xi$-loop at $t=t_i$, the time loop then advances the driver macroparticles from  $t=t_i$ to $t=t_{i+1}$ according to Eq.(\ref{eq:EoM_beam_1},\ref{eq:EoM_beam_2}) using a volume-preserving algorithm (VPA)~\cite{VPA_2015}. The currents and densities of the driver particles are then collected and deposited onto the grid similar to the way it was done for the plasma macroparticles, and they also enter the source terms ${\bf J}$. When a laser driver is present, its envelope equation (\ref{eq:laser_Enve}) is rearranged to a Poisson-like equation and solved using the same MG solver.

\subsection{Multigrid Solver with Nonuniform Grids}\label{sec_Nonuniform}

In addition to the adaptive step size refinement in the $\xi$ direction, WAND-PIC also implemented the nonuniform grids at the transverse plane.  To make sure the convergence of our MG solver, the second-order operators used in Eq.(\ref{eq:Psi_Matrix}) need to be properly approximated on the nonuniform grids.  Figure~\ref{Fig5}~(a) shows a 9-points stencil of nonuniform 2D grids. Consider the general form of wakefield equations: $\nabla^2_\perp\Psi+M\Psi=J$ and apply Taylor expansion around the central point $\Psi_{i,j}$, the 2D Laplacian can be approximated with the following:

\begin{figure}[htp!]
\centering
\includegraphics[width=0.8\columnwidth]{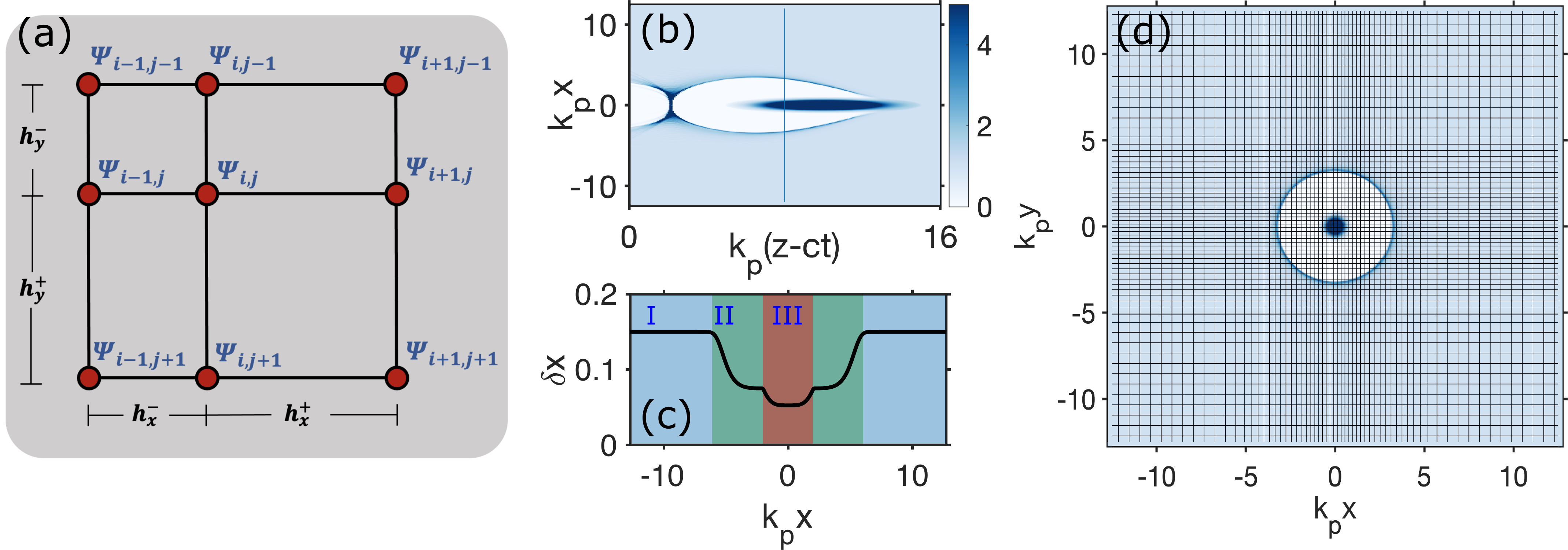}\\
\caption{Nonuniform grids and its application. (a) A 9-points stencil of a nonuniform 2D grids. (b) A common beam-driven simulation. A Gaussian driver bunch is used: $\sigma_x^d=\sigma_y^d=0.5k_p^{-1}$, $\sigma_z^d=2.5k_p^{-1}$, and peak density $=20n_0$. The witness bunch has: $\sigma_x^w=\sigma_y^w=0.4k_p^{-1}$, $\sigma_z^w=0.5k_p^{-1}$, and peak density $=10n_0$ . (c) Grid size $\delta x$ as a function of $x$ (same for the $\delta y$). There are three main regions: (I) plasma region, (II) bubble region, and (III) beam region.  (d) 2D electron density distribution at $k_p(z-ct)=8$, overlaid with the transverse grids. To make the grids visually clear, we enlarged the grid size by 5 times, the actual grids used in simulation are 5 times denser everywhere.}
\label{Fig5}
\end{figure}

\begin{flalign}
\left(\frac{\partial^2}{\partial x^2}+\frac{\partial^2}{\partial y^2}\right)\Psi_{i,j} &= \frac{2}{h_x^-h_x^+h_x^a}\left(h_x^+\Psi_{i-1,j}-h_x^a\Psi_{i,j}+h_x^-\Psi_{i+1,j}\right) + \frac{h_x^b}{3}\frac{\partial^3\Psi_{i,j}}{\partial x^3}\nonumber\\
&+\frac{2}{h_y^-h_y^+h_y^a}\left(h_y^+\Psi_{i,j-1}-h_y^a\Psi_{i,j}+h_y^-\Psi_{i,j+1}\right) + \frac{h_y^b}{3}\frac{\partial^3\Psi_{i,j}}{\partial y^3} + O(h^2),\label{eq:2D_Laplac}
\end{flalign}
where $h$ is the local characteristic grid size. $h_x^a=h_x^++h_x^-$, $h_x^b=h_x^+-h_x^-$, and same for those in the y-direction. In the case of uniform grids: $h_x^-=h_x^+=h_y^-=h_y^+=h$, the Eq.~(\ref{eq:2D_Laplac}) will be reduced to a five-point difference operator: 
\begin{flalign}
\left(\frac{\partial^2}{\partial x^2}+\frac{\partial^2}{\partial y^2}\right)\Psi_{i,j} &= \frac{1}{h^2}\left(\Psi_{i-1,j}+\Psi_{i+1,j}+\Psi_{i,j-1}+\Psi_{i,j+1}-4\Psi_{i,j}\right)+ O(h^2). \label{eq:2D_Laplac_Uni}
\end{flalign}
This 5-point difference scheme is commonly used in codes with 2D uniform grids. However, if we still use the 2nd order approximation for Eq.~(\ref{eq:2D_Laplac}), the additional truncation error $\left(h_x^b\partial^3\Psi_{i,j}/\partial x^3+h_y^b\partial^3\Psi_{i,j}/\partial y^3\right)$ will enter, and that increases with local grid non-uniformity. Therefore, the 3rd-order difference must be kept in Eq.~(\ref{eq:2D_Laplac}), to make sure the convergence of the multigrid solver. Luckily, one can approximate the 3rd-order terms:
\begin{flalign}
h_x^b\frac{\partial^3\Psi_{i,j}}{\partial x^3}+h_y^b\frac{\partial^3\Psi_{i,j}}{\partial y^3}
&=
\left(h_x^b\frac{\partial}{\partial x}+h_y^b\frac{\partial}{\partial y}\right)\left(\frac{\partial^2\Psi_{i,j}}{\partial x^2}+\frac{\partial^2\Psi_{i,j}}{\partial y^2}\right)
-
\left(h_x^b\frac{\partial^3\Psi_{i,j}}{\partial x\partial y^2}+h_y^b\frac{\partial^3\Psi_{i,j}}{\partial y\partial x^2}\right)\nonumber\\
&\approx\left(h_x^b\frac{\partial}{\partial x}+h_y^b\frac{\partial}{\partial y}\right)\left(J_{i,j}-M_{i,j}\Psi_{i,j}\right)
-h_x^b\left[\frac{2}{h_y^-h_y^+h_y^a}\frac{\partial}{\partial x}\left(h_y^+\Psi_{i,j-1}-h_y^a\Psi_{i,j}+h_y^-\Psi_{i,j+1}\right)\right]\nonumber\\
&-h_x^b\left[\frac{2}{h_x^-h_x^+h_x^a}\frac{\partial}{\partial y}\left(h_x^+\Psi_{i-1,j}-h_x^a\Psi_{i,j}+h_x^-\Psi_{i+1,j}\right) \right],
\label{eq:2D_Laplac_3rd}
\end{flalign}
where the 1st-order difference operator can be approximated with 3 points, for example: 
\begin{equation}
\frac{\partial\Psi_{i,j}}{\partial x} = \frac{h_x^-}{h_x^ah_x^+}(\Psi_{i+1,j}-\Psi_{i,j})+\frac{h_x^+}{h_x^ah_x^-}(\Psi_{i,j}-\Psi_{i-1,j}). 
\end{equation}
By plugging Eq.~(\ref{eq:2D_Laplac_3rd}) into Eq.~(\ref{eq:2D_Laplac}), we approximate the transverse Laplacian operator with all 9 points. This numerical scheme is used in the relaxation process of our MG solver, and the inclusion of 3rd-order correction: Eq.~(\ref{eq:2D_Laplac_3rd}) ensures a good converge rate even with a large local grid non-uniformity. The restriction and elongation processes of the MG solver are also slightly modified to take into account the different weights of adjacent cells on a nonuniform grid~\cite{nonuni_16}. This feature of WAND-PIC enables us to efficiently simulate large domains by deploying finer grids only at the area of interest, for example, the bubble region and the region with driver/witness bunches. Combined with the adaptive step size refinement in the longitudinal direction, we have achieved nonuniform grids in all three directions to take care of the structures with different spatial scales. 

Figure~\ref{Fig5}~(b) shows a simple simulation setup in which the nonuniform grids are applied. In this setup, a driver electron bunch drives the bubble and a smaller witness bunch gets accelerated at the back of the bubble. While the vast area outside of the plasma bubble doesn't deserve a high resolution, the bubble boundaries are sharp and thus need to be resolved with denser grids. Furthermore, the near-axis region needs an even higher resolution to accurately model the bunches' phase space evolutions, e.g. emittance and potential hosing instabilities. Therefore, we divide the transverse plane into three regions: (I) the tranquil plasma region that uses coarse grids: $\delta x = \delta y=0.15k_p^{-1}$, (II) the bubble region that uses dense grids: $\delta x = \delta y=0.07\sim0.1k_p^{-1}$, and (III) the beam region that uses more-dense grids: $\delta x = \delta y=0.05k_p^{-1}$. The variation of grid size $\delta x(x)$ can be found in Fig.~\ref{Fig5}~(c) and the visualization of transverse grids are plotted in Fig.~\ref{Fig5}~(d) at position $z-ct=8k_p^{-1}$, together with the electron densities. With the nonuniform grids, WAND-PIC uses 3 times denser grids near the axis and 2 times denser grids for the bubble boundary compared with the largest grid. Overall, $250\times250$ cells are seeded in the transverse plane. If uniform grids are used: $\delta x=\delta y =0.05 k_p^{-1}$, $500\times500$ cells and quadrupled runtime are expected.

\subsection{Benchmarking}

\begin{figure}[htp!]
\centering
\includegraphics[width=0.8\columnwidth]{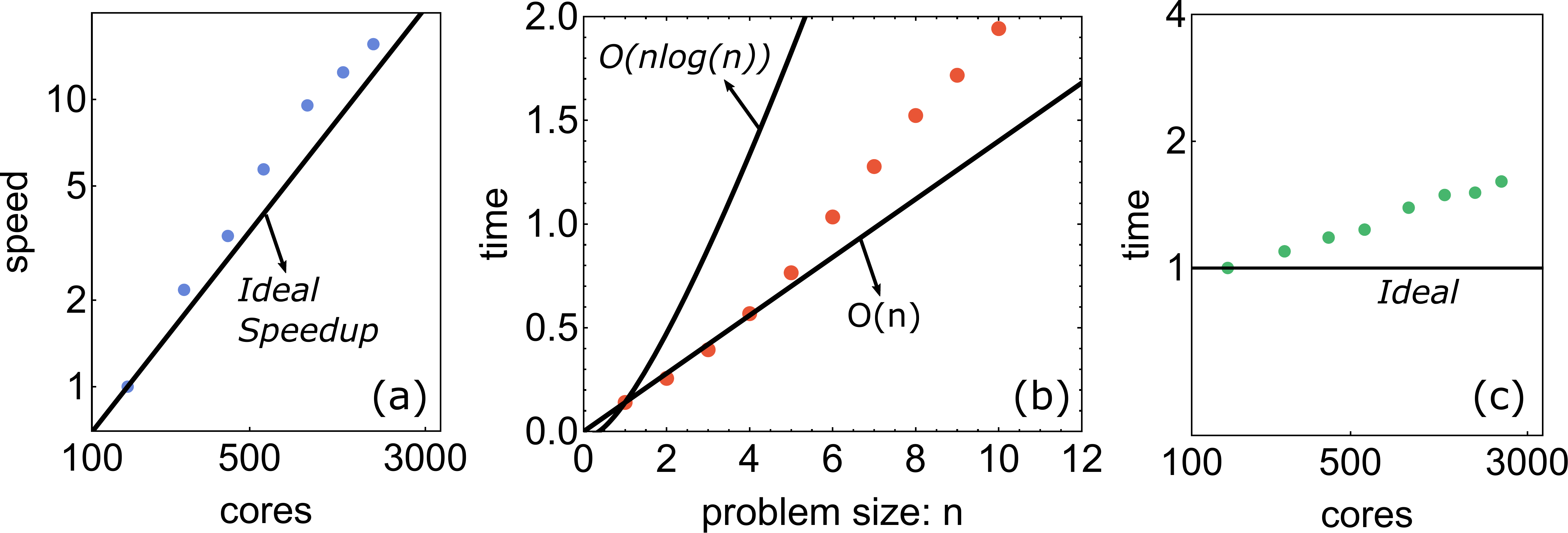}\\
\caption{Benchmarking of WAND-PIC on \textit{Lonestar 5 Architecture (Xeon E5-2690 v3)}.
(a) Strong scaling of WAND-PIC, transverse problem size (number of cells) is fixed at $N_{\rm grid} = 1,000\times 1,000$. (b) The algorithm time complexity of WAND-PIC: the number of cores fixed at $N_{\rm c} = 64$, the problem size $N_{\rm grid} = n \times N_{0}$ is varied ($N_{0} = 320\times 320$). (c) Weak scaling of WAND-PIC for fixed $N_{\rm grid}/N_{\rm c} = 20\times20$.}
\label{Fig6}
\end{figure}

To assess the overall performance of the WAND-PIC under parallelization, benchmarking is conducted and the standard computational quantities, such as strong scaling, weak scaling, and time complexity, were extracted and presented in Fig.~\ref{Fig6}. When taking about the scalings and complexity, the transverse slices are the main subject of study since the $\xi$ direction is not currently parallelized and it has linear time complexity. Note that the scalings and efficiency can be problem-dependent. For example, a linear regime and a blowout regime would generate different loads on the computational cores due to the different movements of trajectories. Therefore, we concentrate on one specific physical scenario: a spherical bubble with a complete blowout driven by an intense laser pulse. We chose the domain size to be three times the bubble size --- the characteristic length of the solution is fixed when we vary computational cores and transverse grids. Laser and domain parameters are presented in Table~\ref{table1}. For this specific physical setup, approximately $80\%$ of the runtime is consumed by the MG solver.

\begin{table}[ht]
\caption{\label{table1}Simulation Parameter}
\begin{ruledtabular}
\begin{tabular}{cccccccc}
\
Domain Size(x-y-z)&$\delta\xi^{(0)}$ \footnote{Adaptive step size is used.} &$\omega_0/\omega_p$ & $a_0$ & $\sigma_x$ & $\sigma_y$ & $\sigma_z$ & Laser Center ($x_0$, $y_0$, $z_0$)\footnote{Center of transverse plane is (0,0), $\xi$ coordinate starts with 0.}\\
\hline
$(25\times25\times25)k_p^{-3}$&$0.08k_p^{-1}$&$20$&$5$&$3.2k_p^{-1}$&$3.2k_p^{-1}$&$2.1k_p^{-1}$&($0$, $0$, $5$)$k_p^{-1}$\\
\end{tabular}
\end{ruledtabular}
\end{table}

Figure~\ref{Fig6}~(a) shows the strong scaling, where the finest transverse grid size (referred to as the problem size) is fixed to be $N_{\rm grid} = 1,000 \times 1,000$ cells. Therefore, $N_{\rm grid}$ determines the transverse resolution. WAND-PIC shows excellent linear speedup up to $N_{\rm c} = 3,000$ cores (note some run-to-run variance due to hardware issues).  The overall time complexity of the WAND-PIC is plotted in Fig.~\ref{Fig6}~(b).  The number of cores was fixed at $N_{\rm c} = 64$, and the normalized problem size $n$ was defined as $n = N_{\rm grid}/N_{0}$, where $N_{0} = 320\times320$ is the base problem size corresponding to $n=1$.  The time complexity of WAND-PIC exhibit a $O(n)$ scaling at a smaller problem size, but deviates from $O(n)$ for larger problem sizes. Overall, time complexity of WAND-PIC falls between $O(n)$ and $O(nlog(n))$. Finally, Fig.~\ref{Fig6}~(c) shows the weak scaling where the problem size per core is fixed:  $N_{\rm grid}/N_{\rm c} = 20\times20$. The parallel efficiency remains $83.3\%$ at $N_{\rm c} = 500$ cores and falls to $62\%$ at $N_{\rm c} = 3,000$ cores. Clearly. the parallel efficiency somewhat decreases when more cores are added, mostly due to increased overheads. Sources of overheads include (i) fast trajectories crossing more sub-domains when the physical size of a sub-domain decreases, thus requiring more send/receive operations; (ii) increased communication between hardware nodes (\textit{Lonestar 5} has $24$ cores per node). Yet, considerable room for improving WAND-PIC still remain, for example, by  implementing load balancing.

\section{Future Code Development and Conclusions}\label{sec:conclusions}
Since the first release of WAND-PIC in 2019~\cite{WAND_PIC}, it has been under continuous improvement. The new features we are developing now include (i) parallelization in the longitudinal dimension through the pipeline technique~\cite{QUICK_2009,HiPACE_2014} which would extend our scalability to hundreds of thousands of cores; (ii) automatic load balancing which would reduce the overheads and improve the efficiency; (iii) transverse local mesh refinement which would improve the level of details in the physical region we are interested in, for example, the back of the bubble and witness beam; (iv) better multigrid cycles and smoothers which would improve the MG solver performance.  These changes will be applied in the near future and be released to the open-source community.

In conclusion, a new quasi-static 3D parallel PIC code: WAND-PIC has been introduced in this work. With the advanced quasi-static equations which are fully explicit and static, wakefields driven by the relativistic beams or laser pulses are solved without using the predictor-corrector method. WAND-PIC has implemented different types of drivers as well as the interactions between the drivers and is able to simulate various scenarios in the plasma-based accelerators. Comparison between the results of WAND-PIC and a 3D full PIC code (VLPL) show that the WAND-PIC is efficient and accurate in modeling the large bubble driven by a large beam charge and the direct laser acceleration of electrons in the bubble. Good parallel scalings and time complexity are achieved by the use of a parallel MG solver and simplified explicit field-solving procedures.

\section{Acknowledgments}
This work was supported by the DOE grant DE-SC0019431. The authors thank the Texas Advanced Computing Center (TACC) at The University of Texas at Austin for providing HPC resources. The authors would like to also thank Dr. Roopendra Singh Rajawat for suggesting the  WAND-PIC abbreviation.


\begin{thebibliography}{10}

\bibitem{WAND_PIC}
    {\em WAND-PIC Code Repository}, \par\url{https://github.com/tianhongg/WAND-PIC}.

\bibitem{LWFA_1}
    V.~Malka, J.~Faure, Y.~A.~Gauduel, E.~Lefebvre, A.~Rousse, and K.~T.~Phuoc,
    "Principles and applications of compact laser–plasma accelerators,"
    {\em Nat. Phys.}, vol.~4, 447 (2008).

\bibitem{LWFA_2}
    E.~Esarey, C.~B.~Schroeder, and W.~P.~Leemans,
    "Physics of laser-driven plasma-based electron accelerators,"
    {\em Rev. Mod. Phys.}, vol.~81, 1229 (2009).

\bibitem{LWFA_3}
    S.~M.~Hooker,
    "Developments in laser-driven plasma accelerators,"
    {\em Nat. Photonics}, vol.~7, 775 (2013).
    
\bibitem{PWFA_1}
	T. Katsouleas,
	"Physical mechanisms in the plasma wake-field accelerator,"
	{\em Phys. Rev. A.}, vol.~33, 2056 (1986).

\bibitem{GeV_0}
    K.~Nakamura, B.~Nagler, C.~Tóth, C.~G.~R.~Geddes, C.~B.~Schroeder, E.~Esarey, S.~M.~Hooker,
    "GeV electron beams from a centimeter-scale channel guided laser wakefield accelerator,"
    {\em Phys. Plasmas}, vol.~14, 056708 (2007).

\bibitem{GeV_1}

    X.~Wang, R.~Zgadzaj, N.~azel, Z.~Li, S.~A.~Yi, X.~Zhang, W.~Henderson, Y.-Y.~Chang, R.~Korzekwa, H.-E.~Tsai, C.-H. ~Pai, H.~Quevedo, G.~Dyer, E.~Gaul, M.~Martinez, A.~C.~Bernstein, T.~Borger, M.~Spinks, M.~Donovan, V.~Khudik, G.~Shvets, T.~Ditmire and M.~C.~Downer,  
    "Quasi-monoenergetic laser-plasma acceleration of electrons to 2 GeV,"
    {\em Nat. Comms.}, vol.~4, 1988 (2013).
\bibitem{GeV_2}
    W.~P.~Leemans, A.~J.~Gonsalves, H.-S. Mao, K.~Nakamura, C.~Benedetti, C.~B.~Schroeder, Cs.~Tóth, J. Daniels, D.~E.~Mittelberger, S.~S.~Bulanov, J.-L.~Vay, C.~G.~R.~Geddes, and E.~Esarey,
    "Multi-GeV electron beams from capillary-discharge-guided subpetawatt laser pulses in the self-trapping regime,"
    {\em Phys. Rev. Lett.}, vol.~113 245002 (2014).

\bibitem{GeV_3}
    H.~T.~Kim, V. B. Pathak, K. H. Pae, A. Lifschitz, F. Sylla, J. H. Shin,  C. Hojbota, S. Ku. Lee, J. H. Sung, H. W. Lee, E. Guillaume, C. Thaury, K. Nakajima, J. Vieira, L. O. Silva, V. Malka and C. H. Nam,
    "Stable multi-GeV electron accelerator driven by waveform-controlled PW laser pulses,"
    {\em Sci. Rep.}, vol.~7, 10203 (2017).

\bibitem{GeV_2b}
    A. J. Gonsalves, K. Nakamura, J. Daniels, C. Benedetti, C. Pieronek, T. C. H. de Raadt, S. Steinke, J. H. Bin, S. S. Bulanov, J. van Tilborg, C. G. R. Geddes, C. B. Schroeder, Cs. Toth, E. Esarey, K. Swanson, L. Fan-Chiang, G. Bagdasarov, N. Bobrova, V. Gasilov, G. Korn, P. Sasorov, and W. P. Leemans,
    "Petawatt Laser Guiding and Electron Beam Acceleration to 8 GeV in a Laser-Heated Capillary Discharge Waveguide,"
    {\em Phys. Rev. Lett.}, vol.~122, 084801 (2019).

\bibitem{PWFA_GeV_1}
	M. J. Hogan, C. D. Barnes, C. E. Clayton, F. J. Decker, S. Deng, P. Emma, C. Huang, R. H. Iverson, D. K. Johnson, C. Joshi, T. Katsouleas, P. Krejcik, W. Lu, K. A. Marsh, W. B. Mori, P. Muggli, C. L. O’Connell, E. Oz, R. H. Siemann, and D. Walz,
	"Multi-GeV Energy Gain in a Plasma-Wakefield Accelerator,"
	{\em Phys. Rev. Lett}, vol.~95, 054802 (2005).

\bibitem{PWFA_GeV_2}
	M. Litos, E. Adli, W. An, C. I. Clarke, C. E. Clayton, S. Corde, J. P. Delahaye, R. J. England, A. S. Fisher, J. Frederico, S. Gessner, S. Z. Green, M. J. Hogan, C. Joshi, W. Lu, K. A. Marsh, W. B. Mori, P. Muggli, N. Vafaei-Najafabadi, D. Walz, G. White, Z. Wu, V. Yakimenko, and G. Yocky,
	"High-efficiency acceleration of an electron beam in a plasma wakefield accelerator,"
	{\em Nature}, vol.~515, 92 (2014).

\bibitem{FACET_II_PWFA}
	C. Joshi, E. Adli, W. An, C.E. Clayton, S. Corde, S. Gessner, M.J. Hogan, M. Litos, W. Lu, K.A. Marsh, and W.B. Mori, "Plasma wakefield acceleration experiments at FACET II," 
	{\em Plasma Phys. Control. Fusion}, vol.~60, 034001 (2018). 

\bibitem{FACET_II_PWFA_II}
	V. Yakimenko, L. Alsberg, E. Bong, G. Bouchard, C. Clarke, C. Emma, S. Green, C. Hast, M. J. Hogan, J. Seabury, N. Lipkowitz, B. O’Shea, D. Storey, G. White, and G. Yocky,
	"FACET-II facility for advanced accelerator experimental tests,"
	{\em Phys. Rev. Accel. Beams}, vol.~22, 101301 (2019).

\bibitem{Tajima_1979}
     T. Tajima, J. M. Dawson, "Laser Electron Accelerator,"
	{\em Phys. Rev. Lett}, vol.~43, 267 (1979).

\bibitem{Joshi_1984}
	C. Joshi, W. B. Mori, T. Katsouleas, J. M. Dawson, J. M. Kindel, and D. W. Forslund,
	"Ultrahigh gradient particle acceleration by intense laser-driven plasma density waves,"
	{\em Nature}, vol.~311, 525 (1984).

\bibitem{Lu_GeV}
    W.~Lu, M. Tzoufras, C. Joshi, F. S. Tsung, W. B. Mori, J. Vieira, R. A. Fonseca, and L. O. Silva,
    "Generating multi-GeV electron bunches using single stage laser wakefield acceleration in a 3D nonlinear regime,"
    {\em Phys. Rev. ST Accel. Beams},  vol.~10, 061301  (2007).

 \bibitem{TR_PisinChen_PRL86}
    P. Chen, J. J. Su, J. M. Dawson, K. L. F. Bane, and P. B. Wilson,
    "Energy Transfer in the Plasma Wake-Field Accelerator,"
    {\em Phys. Rev. Lett.}, vol.~56, 1252 (1986).

\bibitem{blumenfeld_energy_2007}
     I.~Blumenfeld, C. E.~Clayton, F.-J.~Decker, M.~J.~Hogan, C.~Huang, R.~Ischebeck, R.~Iverson, C.~Joshi, T.~Katsouleas, N.~Kirby, W.~Lu, K.~A.~Marsh, W.~B.~Mori, P.~Muggli, E.~Oz, R.~H.~Siemann, D.~Walz, M.~Zhou,
     "Energy doubling of 42 GeV electrons in a metre-scale plasma wakefield accelerator,"
      {\em Nature}, vol.~445, 741--744 (2007).

\bibitem{mourou2011eli_whitebook}
    G. A. Mourou, G. Korn, W. Sandner, and J. L. Collier,
    "ELI WHITEBOOK,"
    {\em THOSS Media GmbH}, (2011).

\bibitem{Korea_4PW}
    J.~H.~Sung, H.~W.~Lee, J.~Y.~Yoo, J.~W.~Yoon, C.~W.~Lee, J.~M.~Yang, Y.~J.~Son, Y.~H.~Jang, S.~K.~Lee, and C.~H.~Nam,
    "4.2 PW, 20 fs Ti: sapphire laser at 0.1 Hz,"
    {\em Opt. Lett.}, vol~42, 11 (2017).

\bibitem{Apollon_10PW}
    B.~Le Garrec, D.~N.~Papadopoulos, C.~Le~Blanc, J.~P.~Zou, G.~Chériaux, P.~Georges, F.~Druon, L.~Martin, L.~Fréneaux, A.~Beluze, N.~Lebas; F.~Mathieu, P.~Audebert,
    "Design update and recent results of the Apollon 10 PW facility,"
    {\em Proc. SPIE}, vol~10238, 80 (2017).

\bibitem{FACET-II-Design}
  FACET-II Technical Design Report No. SLAC-R-1072, (2016).

\bibitem{FACET-II-2020}
  V.~Yakimenko, L.~Alsberg, E.~Bong, G.~Bouchard, C.~Clarke, C.~Emma, S.~Green, C.~Hast, M.~J.~Hogan, J.~Seabury, N.~Lipkowitz, B.~O’Shea, D.~Storey, G.~White, and G. Yocky,
  "FACET-II facility for advanced accelerator experimental tests,"
  {\em Phys. Rev. Accel. Beams} vol.~22, 101301 (2019).

\bibitem{Colliders}
    C. B. Schroeder, E.~Esarey, C. G. R. Geddes, C. Benedetti, and W. P. Leemans,
    "Physics considerations for laser-plasma linear colliders,"
    {\em Phys. Rev. ST Accel. Beams}, vol.~13, 101301 (2010).

\bibitem{PIC_1}
	John M. Dawson,
	"Particle simulation of plasmas,"
	{\em Rev. Mod. Phys.}, vol.~55, 403 (1983).

\bibitem{PIC_2}
	Birdsall, C. K., A. B. Langdon, V. Vehedi, and J. P. Verboncoeur,
	"Plasma Physics via Computer Simulations,"
	Bristol, Adam Hilger, (1991).

\bibitem{YEE}
	K. Yee,
	"Numerical solution of initial boundary value problems involving maxwell's equations in isotropic media,"
	{\em IEEE Trans.}, vol.~14, 302 (1966).

\bibitem{CFL_Condition}	
	R. Courant, K. Friedrichs, and H. Lewy,
	"On the partial difference equations of mathematical physics." IBM journal of Research and Development,"
	{\em  IBM journal of Research and Development}, vol~11, 215 (1967).

\bibitem{WB_JETP_1956}
	A. I. Akhiezer and R. V. Polovin, 
	“Theory of Wave Motion of an Electron Plasma,” 
	{\em Soviet Phys. JETP} vol.~3, 696 (1956).

\bibitem{WB_Dawson_1959}
	John M. Dawson,
	"Nonlinear Electron Oscillations in a Cold Plasma,"
	{\em Phys. Rev.} vol.~113 383 (1959).

\bibitem{WB_Warm_Coffey}
	T. P. Coffey,
	"Breaking of Large Amplitude Plasma Oscillations,"
	{\em Phys. Fluids} vol.~14, 1402 (1971)

\bibitem{WB_Warm_Schroeder}
	C. B. Schroeder, E. Esarey, and B. A. Shadwick,
	"Warm wave breaking of nonlinear plasma waves with arbitrary phase velocities,"
	{\em Phys. Rev. E} vol.~72, 055401 (2005).

\bibitem{WB_Warm_Schroeder2}
	C. B. Schroeder and E. Esarey,
	"Relativistic warm plasma theory of nonlinear laser-driven electron plasma waves,"
	{\em Phys. Rev. E}, vol.~81, 056403 (2010).

\bibitem{Sprangle_1990}
P.~Sprangle, E.~Esarey, and A. Ting, "Nonlinear Theory of Intense Laser-Plasma Interactions,"	
	{\em Phys. Rev. Lett.}, vol.~64, 2011 (1990).

\bibitem{Mora_1996}	
	P. Mora and T. M. Antonsen, Jr.,
	"Electron cavitation and acceleration in the wake of an ultraintense, self-focused laser pulse,"
	{\em Phys. Rev. E}, vol.~53, R2068 (1996).

\bibitem{Mora_1997}			
	P. Mora and T. M. Antonsen Jr.,
	"Kinetic modeling of intense, short laser pulses propagating in tenuous plasmas,"
	{\em Phys. Plasmas}, vol.~4, 217 (1997).

\bibitem{Whittum_1997}
	D. H. Whittum,
	"Transverse two-stream instability of a beam with a Bennett profile,"
	{\rm Phys. Plasmas}, vol.~4, 1154 (1997).

\bibitem{lotov_2003}
	K.~V.~Lotov,
	"Fine wakefield structure in the blowout regime of plasma wakefield accelerators,"
	{\em Phys. Rev. ST Accel. Beams}, vol.~6, 061301 (2003).

\bibitem{lotov_2004}
	K. V. Lotov,
	"Blowout regimes of plasma wakefield acceleration,"
	{\em Phys. Rev. E}, vol.~69, 046405 (2004).

\bibitem{QUICK_2006}
	C. Huang, V.K. Decyk, C. Ren, M. Zhou, W. Lu, W.B. Mori, J.H. Cooley, T.M. Antonsen Jr., T. Katsouleas, "QUICKPIC: A highly efficient particle-in-cell code for modeling wakefield acceleration in plasmas," 	
{\em J. Comp. Phys.}, vol.~217, 658 (2006).

\bibitem{QUICK_2009}		
	B. Feng, C. Huang, V. Decyk, Warren B. Mori, P. Muggli, and T. Katsouleas,
	"Enhancing parallel quasi-static particle-in-cell simulations with a pipelining algorithm."
	{\em J. Comp. Phys.}, vol.~228, 5340 (2009).

\bibitem{QUICK_2013}	
	W. An, V. K. Decyk, W. B. Mori, and T. M. Antonsen Jr,
	"An improved iteration loop for the three dimensional quasi-static particle-in-cell algorithm: QuickPIC,"
	{\em J. Comp. Phys.}, vol.~250 165 (2013).

\bibitem{HiPACE_2014}	
	T. Mehrling, C. Benedetti, C. B. Schroeder, and J. Osterhoff,
	"HiPACE: a quasi-static particle-in-cell code,"
	{\em Plasma Phys. Control. Fusion}, vol.~56, 084012 (2014).

\bibitem{Predictor-Corrector}
	J. C. Butcher,
	"Numerical methods for ordinary differential equations,"
	John Wiley \& Sons, (2016).

\bibitem{rosen_pra91}
	J. B. Rosenzweig, B. Breizman, T. Katsouleas, and J. J. Su,
	"Acceleration and focusing of electrons in two-dimensional nonlinear plasma wake fields,"
    {\em Phys. Rev. A}, vol.~44, R6189 (1991).

\bibitem{pukhov_AppPhysB02}
	A. Pukhov and J. Meyer-ter-Vehn,
	"Laser wake field acceleration: the highly non-linear broken-wave regime," {\em Appl. Phys. B},
    vol.~74, 355 (2002).

\bibitem{My_Driver_2017}
    T.~Wang, V.~Khudik, B.~Breizman, and G.~Shvets,
    "Nonlinear plasma waves driven by short ultrarelativistic electron bunches,"
    {\em Phys. Plasmas}, vol.~24, 103117 (2017).

\bibitem{Hosing_1}
	David H. Whittum, William M. Sharp, Simon S. Yu, Martin Lampe, and Glenn Joyce,
	"Electron-hose instability in the ion-focused regime,"
	{\em Phys. Rev. Lett.}, vol.~67, 991 (1991).

\bibitem{Hosing_2}
	C. Huang, W. Lu, M. Zhou, C. E. Clayton, C. Joshi, W. B. Mori, P. Muggli, S. Deng, E. Oz, T. Katsouleas, M. J. Hogan, I. Blumenfeld, F. J. Decker, R. Ischebeck, R. H. Iverson, N. A. Kirby, and D. Walz,
	"Hosing Instability in the Blow-Out Regime for Plasma-Wakefield Acceleration,"
	{\em Phys. Rev. Lett.}, vol.~99, 255001 (2007).


\bibitem{Pukhov_code}
	A. Pukhov, "Three-dimensional electromagnetic relativistic particle-in-cell code VLPL (Virtual Laser Plasma Lab),"
	{\em J. Plasma Physics}, vol.~61, 425-433 (1999).


\bibitem{My_LEPA_2020}
    T. Wang, V. Khudik, and G. Shvets,
    "Laser-pulse and electron-bunch plasma wakefield accelerator,"
    {\em Phys. Rev. Accel. Beams}, vol.~23, 111304 (2020).

\bibitem{DLA_require_Alex}
    A. V. Arefiev, G. E. Cochran, D. W. Schumacher, A. P.  Robinson, and G. Chen,
    "Temporal resolution criterion for correctly simulating relativistic electron motion in a high-intensity laser field,"
    {\em Phys. Plasmas}, vol.~22, 013103 (2015).

\bibitem{Multigrid_1}
	W. L. Briggs, V. E. Henson, and S. F. McCormick,
	"A multigrid tutorial,"
	Society for Industrial and Applied Mathematics, (2000).

\bibitem{Multigrid_2}
	E. Chow, R. D. Falgout, J. J. Hu, R. S. Tuminaro, and U. Meier-Yang,
	"A survey of parallelization techniques for multigrid solvers, in Parallel Processing for Scientific Computing,"
	M. A. Heroux, P. Raghavan, and H. D. Simon, eds., Software, Environments, and Tools 20, SIAM, Philadelphia, 2006, pp. 179–201.

\bibitem{morshed_pop10}
	S. Morshed, T. M. Antonsen, and J. P. Palastro,
	"Efficient simulation of electron trapping in laser and plasma wakefield acceleration,"
	{\em Phys. Plasmas}, vol.~17, 063106 (2010).

\bibitem{jain_pop15}
	Neeraj Jain, John Palastro, T. M. Antonsen, Jr., Warren B. Mori, and Weiming An,
	"Plasma wakefield acceleration studies using the quasi-static code WAKE,"
	{\em Phys. Plasmas}, vol.~22, 023103 (2015).

\bibitem{QS_Pulse}
	E. Esarey, P. Sprangle, J. Krall, A. Ting,
 	"Self-focusing and guiding of short laser pulses in ionizing gases and plasmas,"
 	{\em IEEE J. Quantum Electron.},  vol.~33, 1879 (1997)


\bibitem{Xi_prl}
        Xi~Zhang, V.~N.~Khudik, and G.~Shvets,
        "Synergistic Laser-Wakefield and Direct-Laser Acceleration in the Plasma-Bubble Regime,"
        {\em Phys. Rev. Lett.}, vol.~114, 184801 (2015).

\bibitem{zhang_ppcf}
        X. Zhang, V. N. Khudik, A. Pukhov and G. Shvets,
        "Laser wakefield and direct acceleration with ionization injection,"
        {\em Plasma Phys. Control. Fusion}, vol.~58, 034011 (2016).

\bibitem{shaw_ppcf}
        J.~L.~Shaw, F.~S.~Tsung, N.~Vafaei-Najafabadi, K.~A.~Marsh, N.~Lemos, W.~B.~Mori and C.~Joshi,
        "Role of direct laser acceleration in energy gained by electrons in a laser wakefield accelerator with ionization injection,"
        {\em Plasma Phys. Control. Fusion}, vol.~56, 084006 (2017).
\bibitem{shaw_ppcf2}
        J.~L.~Shaw, N.~Lemos, K.~A.~Marsh, D.~H.~Froula and C.~Joshi,
        "Experimental signatures of direct-laser-acceleration-assisted laser wakefield acceleration,"
        {\em Plasma Phys. Control. Fusion}, vol.~60, 044012 (2018).

\bibitem{zhang_ppcf_2}
        X.~Zhang, T.~Wang, V.~N.~Khudik, A.~C.~Bernstein, M.~C.~Downer, and G.~Shvets,
        "Effects of laser polarization and wavelength on hybrid laser wakefield and direct acceleration,"
        {\em Plasma Phys. Control. Fusion}, vol.~60, 105002 (2018).

\bibitem{Kh_2018}
        V.~N.~Khudik, Xi~Zhang, T.~Wang, and G.~Shvets,
        "Far-field constant-gradient laser accelerator of electrons in an ion channel,"
        {\em Phys. Plasmas}, vol.~25, 083101 (2018).

\bibitem{My_DLA_2019}
    T. Wang, V. Khudik, A. Arefiev, and G. Shvets,
    "Direct laser acceleration of electrons in the plasma bubble by tightly focused laser pulses,"
    {\em Phys. Plasmas}, vol.~26, 083101 (2019).

\bibitem{stupakov_2016}
    G.~Stupakov, B.~Breizman, V.~Khudik, and G.~Shvets,
    "Wake excited in plasma by an ultrarelativistic pointlike bunch,"
    {\em Phys. Rev. Accel. Beams}, vol.~19, 101302 (2016).

\bibitem{Lu-Beam-Theory}
    W.~Lu, C.~Huang, M.~Zhou, M.~Tzoufras, F.~S.~Tsung, W.~B.~Mori, T.~Katsouleas, "A nonlinear theory for multidimensional relativistic plasma wave wakefields,"
    {\em Phys. Plasmas}  vol.~13, 056709 (2006).


\bibitem{Injection_Theory}
  I. Yu. Kostyukov, E.N. Nerush, A. Pukhov, and V. Seredov,
  ”Electron Self-Injection in Multidimensional Relativistic-Plasma Wake Fields,”
  {\em Phys. Rev. Lett.}, vol.~103, 175003 (2009).


\bibitem{Khudik_2016}
        Khudik V N, Arefiev A V, Zhang X and Shvets G,
        "Universal scalings for laser acceleration of electrons in ion channels,"
        {\em Phys. Plasmas}, vol.~23, 103108 (2016).

\bibitem{Krash_2018}
        Y.~Zhang and S.~I.~Krasheninnikov,
        "Electron dynamics in the laser and quasi-static electric and magnetic fields,"
        {\em Phys. Lett. A}, vol.~382, 1801 (2018).

\bibitem{pukhov2002_DLA}
	A. Pukhov,
	"Strong field interaction of laser radiation,"
	{\em Rep. Prog. Phys.},  vol.~66, 47 (2003).

\bibitem{CEP_observable}
        J. Huijts, I. Andriyash, L. Rovige, A. Vernier, and J. Faure,
        "Identifying observable carrier-envelope phase effects in laser wakefield acceleration with near-single-cycle pulses", {\em Phys.Plasmas }, {\bf 28}, 043101 (2021).
\bibitem{Zhengyan}
        S. Xu, J. Zhang, N. Tang, S. Wang, W. Lu, and Z. Li,
        "Periodic self-injection of electrons in a fewcycle laser driven oscillating plasma wake",
        {\em AIP Advances}, {\bf 10},095310,(2020)

\bibitem{Salehi}
        F. Salehi, M. Le, L. Railing, M. Kolesik, and H. M. Milchberg,
        "Laser-accelerated, low divergence 15 MeV quasi-monoenergetic electron bunches at 1 kHz",
        {\em Phys. Rev. X  } {\bf 11}, 021055 (2021).

\bibitem{Jihoon}
        J. Kim, T. Wang, V. Khudik, and G. Shvets,
       "Subfemtosecond wakefield injector and accelerator based on an undulating plasma bubble controlled by a laser phase", {\em Phys. Rev. Lett} {\bf 127}, 164801 (2021).

\bibitem{kost_cep}
        E. N. Nerush, and I. Yu. Kostyukov,
        "Carrier-Envelope Phase Effects in Plasma-Based Electron Acceleration with Few-Cycle Laser Pulses",
        {\em Phys. Rev. Lett.}, {\bf 103}, 035001 (2009).

\bibitem{Jihoon2}
		J. Kim, T. Wang, V. Khudik, and G. Shvets, 
		"Polarization control of electron injection and acceleration in the plasma by a self steepening laser pulse", arXiv:2111.03014 (2021).

\bibitem{Boris_pusher}
		J. Boris, in "Proceedings of the Fourth Conference on Numerical Simulation of Plasmas",
		{\em PNaval Research Laboratory, Washington DC} 3 (1970)

\bibitem{VPA_2015}
	R. Zhang, J. Liu, H. Qin,   Y. Wang, Y. He, and Y. Sun,
	"Volume-preserving algorithm for secular relativistic dynamics of charged particles,"
	{\em Phys. Plasmas}, vol.~22, 044501 (2015).

\bibitem{nonuni_16}
	F. Ghaffar, N. Badshah, S. Islam, and M. A. Khan,
	"Multigrid method based on transformation-free high-order scheme for solving 2D Helmholtz equation on nonuniform grids,"
	{\em Adv Differ Equ.},vol.~2016, 19 (2016)


\end{thebibliography}
 \end{document}